\documentclass[journal,twoside,web]{ieeecolor}

\usepackage{soul,color}
\usepackage{algorithm}
\usepackage{algpseudocode}
\usepackage{graphicx}
\usepackage{epstopdf}
\usepackage{generic}
\usepackage{cite}
\usepackage{amsmath,amssymb,amsfonts}
\usepackage{graphicx}
\usepackage{textcomp}

\usepackage{titlesec}
\usepackage{subfigure}
\let\labelindent\relax
\usepackage[shortlabels]{enumitem}

\makeatletter
\let\NAT@parse\undefined
\makeatother
\usepackage[colorlinks,citecolor=green,urlcolor=blue,bookmarks=false,hypertexnames=true]{hyperref}  

\def\BibTeX{{\rm B\kern-.05em{\sc i\kern-.025em b}\kern-.08emhttps://www.overleaf.com/project/620a064f44919ae89c264866
    T\kern-.1667em\lower.7ex\hbox{E}\kern-.125emX}}
\begin{document}
\title{Sparsity Based Non-Contact Vital Signs Monitoring of Multiple People Via FMCW Radar}
\author{Yonathan Eder, \IEEEmembership{Graduate Student Member, IEEE}, and Yonina C. Eldar, \IEEEmembership{Fellow, IEEE}
\thanks{
The authors are with the Faculty of Mathematics and Computer Science, Weizmann Institute of Science, Rehovot,
Israel (e-mail: yoni.eder@weizmann.ac.il, yonina.eldar@weizmann.ac.il).}   }

\maketitle

\begin{abstract}
Non-contact technology for monitoring multiple people's vital signs, such as respiration and heartbeat, has been investigated in recent years due to the rising cardiopulmonary morbidity, the risk of transmitting diseases, and the heavy burden on the medical staff. Frequency modulated continuous wave (FMCW) radars have shown great promise in meeting these needs. However, contemporary techniques for non-contact vital signs monitoring (NCVSM) via FMCW radars, are based on simplistic models, and present difficulties coping with noisy environments containing multiple objects. In this work, we develop an extended model of FMCW radar signals in a noisy setting containing multiple people and clutter. By utilizing the sparse nature of the modeled signals in conjunction with human-typical cardiopulmonary features, we can accurately localize humans and reliably monitor their vital signs, using only a single channel and a single-input-single-output setup. To this end, we first show that spatial sparsity allows for both accurate detection of multiple people and computationally efficient extraction of their Doppler samples, using a joint sparse recovery approach. Given the extracted samples, we develop a method named Vital Signs based Dictionary Recovery (VSDR), which uses a dictionary-based approach to search for the desired rates of respiration and heartbeat over high-resolution grids corresponding to normal cardiopulmonary activity. The advantages of the proposed method are illustrated through examples that combine the proposed model with real data of $30$ monitored individuals. We demonstrate accurate human localization in a clutter-rich scenario that includes both static and vibrating objects, and show that our VSDR approach outperforms existing techniques, based on several statistical metrics. The findings support the widespread use of FMCW radars with the proposed algorithms in healthcare.
\end{abstract}

\section{Introduction}
\label{sec:intro}
\IEEEPARstart{I}{n} the last decade, the rise in chronic health conditions alongside the increase in the elderly population, have resulted in a growing need for health care approaches that emphasize long-term monitoring in addition to urgent intervention  \cite{crossley2011connect,brownsell1999future,boric2002wireless,nangalia2010health,lin1999applying}. Monitoring of human vital signs, however, entails numerous difficulties. First, current monitoring devices are typically in physical contact with the measured body, therefore may lead to irritation or general discomfort to the patient, and can be easily detached. Second, usually monitoring devices are connected to patients by medical staff, whether in clinics or hospitals, by a time-consuming interaction that increases the risk of infections and disease transmission, especially during times of pandemics, such as COVID-19. In addition, the manner of connection greatly affects the results, thus it requires considerable skill and experience. Beyond that, many medical teams suffer from high workloads which ultimately lead to an increase in mortality, infections and duration of hospitalization, e.g. in intensive care units \cite{ceballos2015psychosocial,hillhouse1997investigating,greenglass2001workload, almenyan2021effect}.
 
Remote sensing technology such as radar systems can be ideal in these situations since they do not require users to wear, carry, or interact with any additional electronic device \cite{fioranelli2019radar}. In recent years, several works addressed this issue, attempting to remotely monitor human vital signs such as respiration rate (RR) and heart rate (HR), using radars {\cite{xiao2006ka, gu2012accurate, gu2019noncontact,zhao2018noncontact, park2007arctangent,lai2010wireless,vodai2021enhancement,wang2014hybrid, ahmad2018vital,mercuri2019vital,sacco2020fmcw,adib2015smart, lee2019novel,alizadeh2019remote,antolinos2020cardiopulmonary, wang2020remote,turppa2020vital,kim2018low}}. Initially, the family of continuous wave (CW) radars was proposed as simple and reliable devices for remote measurement of cardiac-related chest movements and respiratory activity \cite{ xiao2006ka, gu2012accurate, gu2019noncontact,zhao2018noncontact, park2007arctangent }, having the advantages of low transmission power and high sensitivity. However, they do not provide patient distance information from the radar nor can they separate returns from different objects. To overcome this limitation, many works have turned to using frequency modulated continuous wave (FMCW) radars {\cite{vodai2021enhancement,wang2014hybrid, ahmad2018vital,mercuri2019vital,sacco2020fmcw,adib2015smart, lee2019novel,alizadeh2019remote,antolinos2020cardiopulmonary, wang2020remote,turppa2020vital,kim2018low}}. FMCW technology allows spatial separation and potentially monitoring of several people simultaneously, which can reduce both loads and financial costs. However, accurate simultaneous extraction of multiple-people's cardiopulmonary activity using FMCW radars, is still a challenge in terms of performance, and currently lacks adequate mathematical modelling, leading to sub-optimal solutions.

\begin{figure*}[htbp!]  
\begin{center}
{\includegraphics[width=1\textwidth]{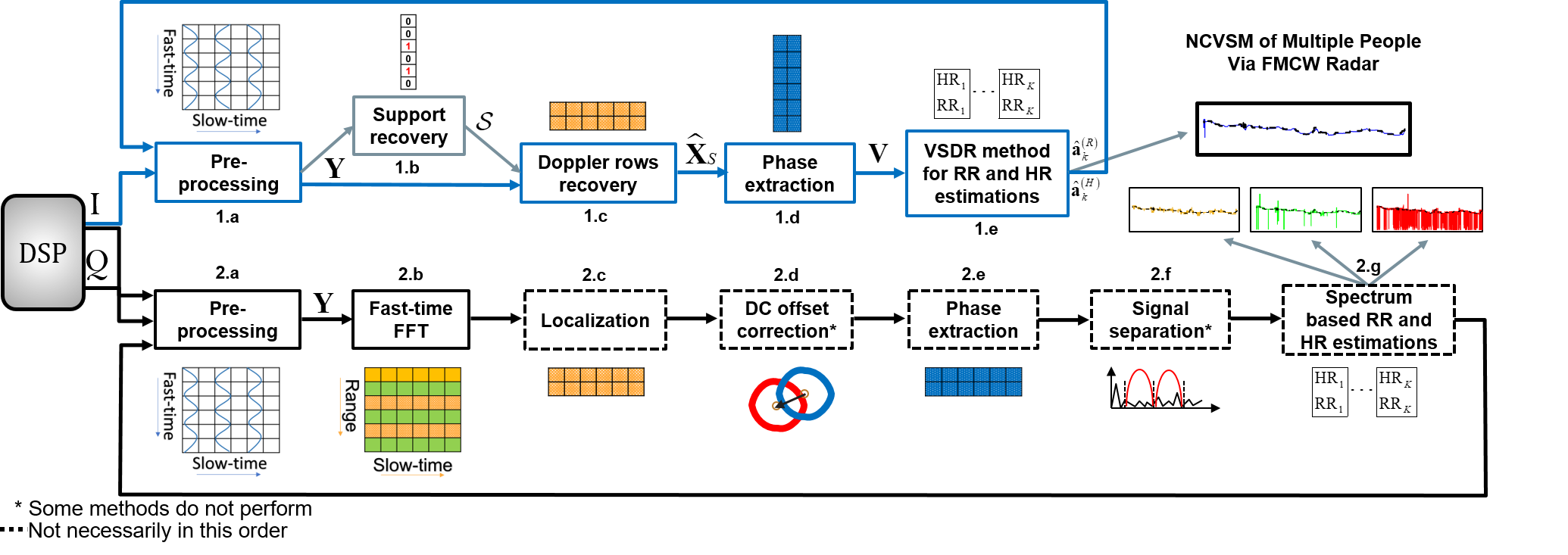}}
\end{center} 
\caption{Block diagram of NCVSM of multiple people via FMCW radar.  \textbf{(Top)} The proposed approach: yields accurate estimates of RR and HR by exploiting the sparse composition of the input data in accordance with human cardiopulmonary properties, using only a single channel. \textbf{(Bottom)} The conventional framework: lacks adequate theoretical explanations and presents difficulties in dealing with noisy, cluttered environments.}  
\label{block_diagram}
\end{figure*}

The conventional algorithmic framework for non-contact vital signs monitoring (NCVSM) of multiple people using FMCW radars, is a recurring process in which an estimate of human vital signs is evaluated and recorded at each fixed time interval, as illustrated in the bottom diagram of Fig. \ref{block_diagram}. In each iteration, a processing matrix is compiled from the raw data obtained by both the \textit{In-phase} (I) and \textit{Quadrature} (Q) channels, through which the samples attributed to human cardiopulmonary activity are located and extracted. Given estimated human thoracic vibrations, various techniques can be used to monitor each human's desired RR and HR.

Traditionally, at each monitoring repetition both the I and Q channels are used to assemble a convenient complex matrix for further processing {\cite{lai2010wireless,vodai2021enhancement,wang2014hybrid, ahmad2018vital,mercuri2019vital,sacco2020fmcw,adib2015smart, lee2019novel,alizadeh2019remote,antolinos2020cardiopulmonary, wang2020remote,turppa2020vital,kim2018low}}. The main drawbacks of using both the I and Q channels, are the lack of perfect orthogonality, and difference in gain levels, in each channel, namely the I/Q imbalance limitation \cite{roudas2007compensation, mahendra2020compensation, tubbax2003compensation},
which may corrupt the desired information to be extracted. This imbalance can be compensated for by methods such as the Gram–Schmidt orthogonalization procedure (GSOP) \cite{fatadin2008compensation}, but there is no guarantee for optimal corrections in realistic noisy environments.

Once the complex map is assembled, a human localization procedure is performed based on a transformed version of the map. Mercuri {\it{et al.}} \cite{mercuri2019vital} performed manual localization based on the intensities of the map, knowing in advance the number of people to be monitored, and their true distance from the radar. Alizadeh {\it{et al.}} \cite{alizadeh2019remote} selected the range bin with the maximal average power. In real-world scenarios, information about the number of people is not typically available, thus relying on spectral magnitudes solely, can produce erroneous decisions due to strong signal reflections obtained from static objects in the radar's field of view (FOV). 

Several works tried to circumvent the effect of clutters on human localization for NCVSM. Adib {\it{et al.}} \cite{adib2015smart} subtracted consecutive time measurements to eliminate reflections off static objects. In \cite{sacco2020fmcw}, the amplitude of a padded fast Fourier transform (FFT) collected over one frame period was compared to the standard deviation based \textit{std} estimate, in a room containing furniture. Antolinos {\it{et al.}} \cite{antolinos2020cardiopulmonary} performed a clipping procedure on a zero-padded FFT map to isolate the target from interfering objects. However, these methods lack adequate theoretical explanations and may be sensitive to vibrating clutters, such as fans.

Once the human-related vectors have been correctly located, the thoracic vibration pattern of each individual is extracted. The most commonly used methods to estimate the considered vital signs from the extracted pattern are based on the discrete Fourier transform (DFT) spectrum, utilizing the property that in resting state, the frequency bands of heartbeat and respiration do not overlap \cite{alizadeh2019remote,sacco2020fmcw,mercuri2019vital,antolinos2020cardiopulmonary}. The latter, and additional DFT-based techniques for estimating the desired rates \cite{turppa2020vital,adib2015smart,kim2018low} are discussed in detail in Subsection \ref{standard_processing}. Despite all the well-known benefits of DFT-spectrum analysis, it presents limitations for the considered problem in terms of both resolution and signal representation which ultimately impairs estimation performance.

In this paper, we develop an extended mathematical signal model for the problem of NCVSM of multiple people using FMCW radars, in a clutter-rich scenario. Our theoretical approach employs only a single channel, and a single-input-single-output (SISO) configuration. In addition to performance amelioration, by doing so, we reduce processing times and circumvent the need to deal with issues related to combining the two channels. Based on the developed model, we propose a complete methodology for human localization and accurate monitoring of their vital signs by leveraging prior knowledge of the FMCW signal structure, for the given problem.

First, the proposed localization technique utilizes a frequency-based understanding of human cardiopulmonary activity, as well as the sparse nature of the data via a joint sparse recovery (JSR) mechanism \cite{rossman2019rapid,eldar2015sampling,eldar2012compressed}. This approach allows for computationally efficient extraction of the relevant Doppler samples throughout the complete monitoring process. Then, by using a Vital Signs based Dictionary Recovery (VSDR) approach, which effectively conducts a frequency search over a dense grid, we exhibit high-resolution NCVSM of multiple people, given their extracted thoracic vibrations. The performance of the proposed methodology is verified through simulations that incorporate synthetic signals based on the developed model with \textit{in vivo} data of $30$ monitored individuals from \cite{schellenberger2020dataset}. This study demonstrates both precise human localization in a multiple object scenario and superior accuracy results for RR and HR monitoring, when compared to state-of-the-art techniques using several statistical metrics.

The rest of the paper is organized as follows. In Section \ref{Pre}, we present the principles of FMCW radar and formulate the problem of single human vital signs monitoring. We then review conventional processing approaches based on this model. In Section \ref{Proposed}, we present an extended FMCW signal model for the case of multiple people monitoring, and develop the proposed methodology. We evaluate the performance of the proposed algorithms and compare them to existing techniques in Section \ref{Numerical}. Finally, Section \ref{Conc} summarizes the main points of this work.

Throughout the paper, we use the following notation. Scalars are denoted by lowercase letters $(a)$, vectors by boldface lowercase letters $(\bf{a})$, sets are given by calligraphic font ($\cal{S}$) and matrices are denoted by boldface capital letters $(\bf{A})$. The $(i,j)$'th element of a matrix $\bf{A}$ is written as ${\bf{A}}(i,j)$, and ${\bf{a}}_l$ is the $l$'th column of ${\bf{A}}$. The notations $(\cdot)^T$, and $(\cdot)^{H}$ indicate the transpose and Hermitian operations, respectively.

\section{Existing FMCW Modeling and Techniques}
\label{Pre}
In this section, we provide the standard 2-D FMCW model for single human RR and HR monitoring at a given time window, based on previous works. We proceed by reviewing existing processing methods that employ this model. An extended representation for multiple people and clutter is presented in Section \ref{Proposed}, through which the proposed NCVSM method is developed.
\subsection{Standard FMCW Model for Single Human NCVSM}
\label{FMCW_principles}
\begin{figure}[t!]
    \centering
    \includegraphics[width=0.5\textwidth]{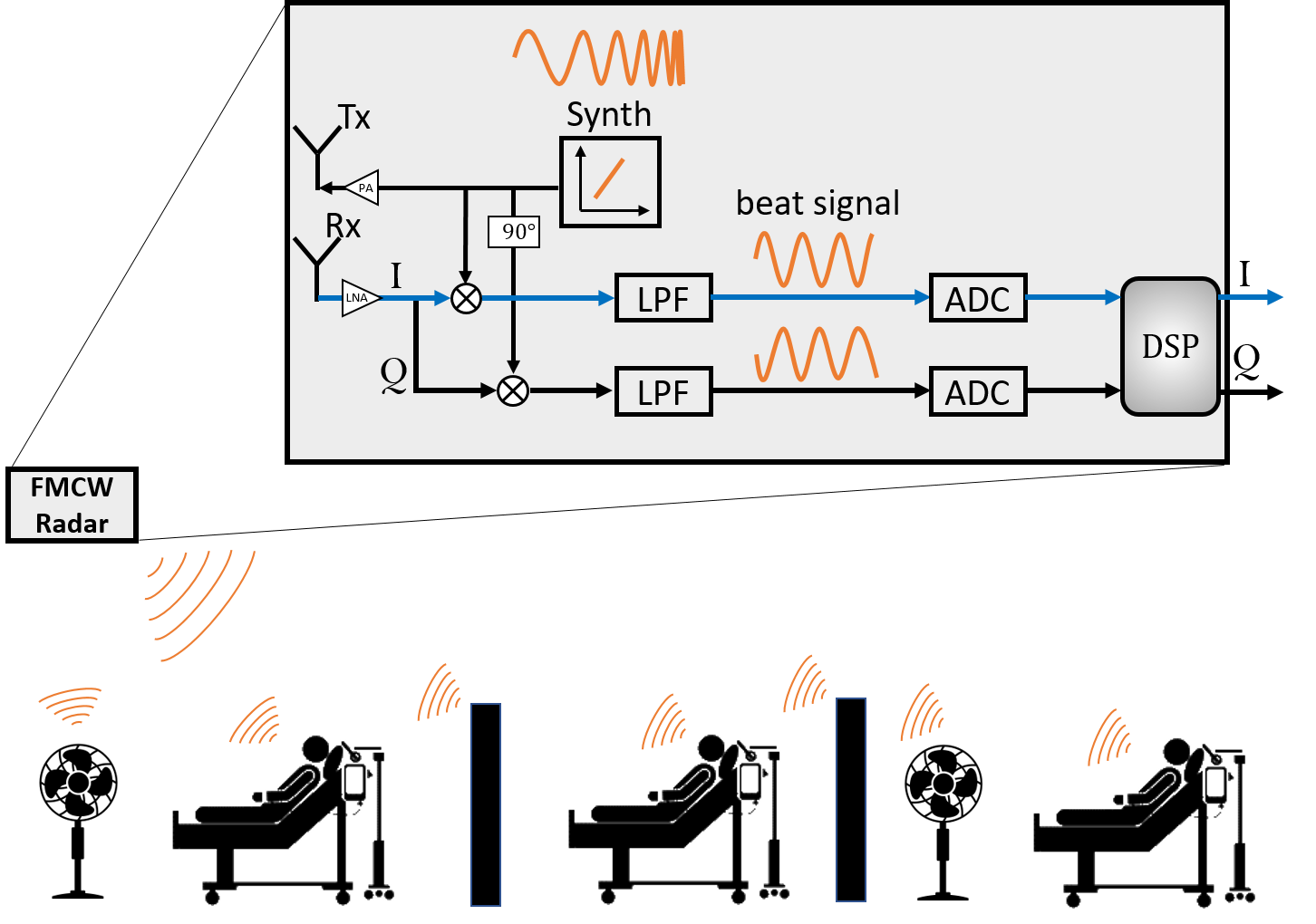}
    \caption{A schematic illustration of FMCW radar's main components and the examined multiple objects scenario. In this study, we show that we can accurately monitor the vital signs of multiple people in a cluttered environment, while using only a single channel  (marked in blue) and a SISO configuration.}
    \label{fig:FMCW_radar_MP}
\end{figure}

\begin{figure}[t!]

    \hspace{-0.6cm}\includegraphics[width=0.54\textwidth]{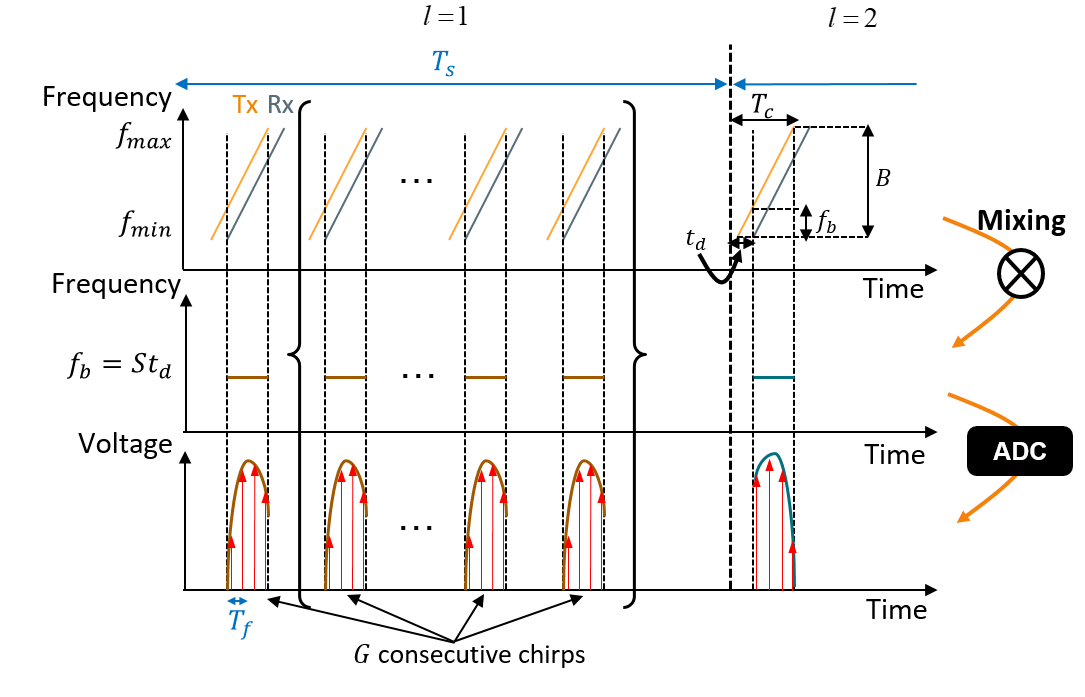}
    \caption{Transmitted and received chirp sequences at each frame. The brackets illustrate the case of transmitting $G$ consecutive chirps per frame, instead of a single one. \textbf{(Top)} Frequency sawtooth waveform. \textbf{(Middle)} Analog beat signal after mixing. \textbf{(Bottom)} Discrete beat signal after ADC sampling.}
    \label{fig:saw-tooth}
\end{figure}

As shown in both figures \ref{fig:FMCW_radar_MP} and \ref{fig:saw-tooth}, a typical FMCW radar transmits a series of signals at a given time frame, called chirps \cite{TIdatasheet}, whose instantaneous frequency linearly increases over time, forming a saw-tooth waveform. The reflected echo signals split between the I/Q channels through which they are mixed with versions of the transmitted one followed by a low-pass-filter (LPF) to obtain analog base-band signals, known as the intermediate-frequency (IF) signals \cite{liu2022radar,wang2020remote}, which are also called beat signals \cite{alizadeh2019remote, turppa2020vital}, as we use in this work. The beat signals in each frame are sequentially sampled by the ADC, resulting in discrete base-band signals corresponding to each channel, which are sent from the radar to a local computer for further processing. 

Suppose a static human target is located $d_0$ meters from the radar, with the antenna facing its thorax. The phenomena of respiration and heartbeat produce small time-varying changes in its relative position, so that its actual radial distance from the radar is 
\begin{equation}
\label{d(t)}
d(t) = d_0+v(t),  
\end{equation}
where $v(t) \ll d_0 $ denotes a human thoracic vibration due to its cardiopulmonary activity. After hitting a human thorax, a transmitted chirp is reflected back to the radar receiver and appears as an attenuated and shifted version of the transmitted one. As illustrated in Fig. \ref{fig:saw-tooth}, the shifting is expressed by the round-trip delay between the radar and the human object, which is approximated by (\ref{d(t)}) to $t_d \approx 2d_0/c$ with $c$ denoting the speed of light.

Relying on derivations from \cite{alizadeh2019remote,antolinos2020cardiopulmonary,wang2020remote} for the FMCW signal model, the continuous beat signal of the \textit{In-phase} channel for a single chirp at a given frame, can be described as 
\begin{equation}
\label{cont_beat_signal}
s_b\left( t \right)\triangleq {x_b}\cos \left( {2\pi {f_b}t + {\psi _b}\left( t \right)} \right),\hspace{0.2cm}t \in [t_d\hspace{0.2cm}T_c],
\end{equation}
with $x_b$, $f_b$ and $\psi_b(t)$ respectively denoting the amplitude, frequency and phase terms of the beat signal, due to the mixing process between the received and transmitted signals in the overlapping time interval $[t_d \hspace{0.2cm}T_c]$, where $T_c$ is the duration of a single chirp, as illustrated in Fig. \ref{fig:saw-tooth}.

The constant beat frequency is defined as
\begin{equation}
\label{beat_freq}
f_b \triangleq  S t_d = \frac{{2S}}{c}{d_0},
\end{equation}
where $S \triangleq B/T_c$ corresponds to the rate of the frequency sweep with $B$ being the chirp's total bandwidth. The time-varying beat phase is
\begin{equation}
\label{psi_t}
{\psi _b}\left( t \right) \triangleq \frac{{4\pi }}{{{\lambda _{\max }}}}\left( {{d_0} + v\left( t \right)} \right),\hspace{0.2cm}t \in [t_d\hspace{0.2cm}T_c],
\end{equation}
where ${{{\lambda _{\max }}}}$ denotes the maximal wavelength of the chirp. In practice, thoracic displacement is approximately constant w.r.t. a chirp's duration \cite{alizadeh2019remote}, hence, in order to extract the temporal variation of a human thorax, consecutive chirps should be transmitted at intervals of $T_s \gg T_c$ seconds, denoted as the \textit{slow-time} sampling interval of $v(t)$. 

Based on (\ref{psi_t}) and the above argument, a discrete phase signal over $L$ given frames can be defined as
\begin{equation}
\label{psi_l}
{\psi _b}\left[ l \right] \triangleq \frac{{4\pi }}{{{\lambda _{\max }}}}\left( {{d_0} + v\left[ l \right]} \right),\hspace{0.2cm} l = 1,...,L,
\end{equation}
where $v\left[ l \right]\triangleq v\left( {l{T_s}} \right)$. Then, the continuous beat signal at each frame $l$, denoted by $\tilde{s}_b\left( t,lT_s \right)={x_b}\cos \left( {2\pi {f_b}t + {\psi _b}\left[ l \right]} \right)$, is sampled by the ADC component at the sampling interval $T_f$, considered as the \textit{fast-time} sampling period of $\tilde{s}_b\left( t,lT_s \right)$ (Fig. \ref{fig:saw-tooth}, bottom). This yields the following 2-D discrete beat signal of the \textit{In-phase} channel
\begin{equation}
\label{2D_discrete_beat_cos}
y_{I}\left[ {n,l} \right] \triangleq \tilde{s}_b\left( {n{T_f},l{T_s}} \right) = x_b\cos \left( {2\pi {f_b}n{T_f} + {\psi _b}\left[ l \right]} \right),
\end{equation}
where $n=1,...,N$ and $l=1,...,L$. The discrete beat signal obtained by the parallel \textit{Quadrature} channel, which is a $90^{\circ}$ shifted version of (\ref{2D_discrete_beat_cos}), is used to compose the following complex exponential term
\begin{equation}
\label{2D_discrete_beat_exp}
y\left[ {n,l} \right] \triangleq \tilde{x}_b\exp \left( {j\left( {2\pi {f_b}n{T_f} + {\psi _b}\left[ l \right]} \right)} \right),\quad {\begin{cases} n=1,...,N \\
                     l=1,...,L
       \end{cases}}.
\end{equation}

The samples in (\ref{2D_discrete_beat_exp}) form a $2$-D measurement matrix ${\bf{Y}} \in \mathbb{C}^{N\times{L}}$ where ${\bf{Y}}(n,l)=y\left[ {n,l} \right]$. The samples along the row dimension are referred to as the \textit{fast-time} samples and are related to the distance of the human object $d_0$ for the $l$'th frame via (\ref{beat_freq}).
The samples along the column dimension of (\ref{2D_discrete_beat_exp}) are referred to as the \textit{slow-time} samples and are associated with the human thoracic vibration function $v\left[ l \right]$, which is reflected in varying phase values between successive frames (\ref{psi_l}). Thus, by estimating $f_b$ and $\{\psi _b\left[ l \right]\}_{l=1}^L$, it is possible to respectively evaluate $d_0$ and $\{v\left[ l \right]\}_{l=1}^L$ by relations (\ref{beat_freq}) and (\ref{psi_l}), from which various methods can be used to extract the corresponding RR and HR at the given time window.
\subsection{Existing Techniques for NCVSM Via FMCW Radar}
\label{standard_processing}
With the help of the lower diagram of Fig. \ref{block_diagram}, in the following we review existing processing techniques for NCVSM based on the standard FMCW model presented in (\ref{2D_discrete_beat_exp}). 

First, the raw signals received from the I and Q channels are pre-processed to construct ${\bf{Y}}$ (Fig. \ref{block_diagram}, block $2$.a). Next, to locate the humans in the radar's FOV and their Doppler information, the \textit{Range vs. Slow-Time} map \cite{alizadeh2019remote} is computed by performing a row-wise \textit{fast-time} FFT over the complex representation of ${\bf{Y}}$, and converting the \textit{fast-time} axis to a distance-based one, using (\ref{beat_freq}) (Fig. \ref{block_diagram}, block $2$.b).

Two widely used spectrum-based methods for identifying human-related range bins from the given map (Fig. \ref{block_diagram}, block $2$.c), are the \textit{manual localization} \cite{mercuri2019vital} and the \textit{Maximum Average Power} \cite{alizadeh2019remote} methods. 
The motivation for exploiting the map's magnitudes stems from (\ref{beat_freq}) and (\ref{2D_discrete_beat_exp}), but since the model in (\ref{2D_discrete_beat_exp}) does not address the presence of clutters, it can lead to a selection of high-intensity disturbances instead of proper localization of the people to be monitored. Sacco {\it{et al.}} \cite{sacco2020fmcw} used the \textit{std} estimate to distinguish between human and static objects. However, since \textit{std} does not exploit characteristics of human thoracic motion, a general search for vibrations may result in an incorrect selection of other vibrating objects, such as fans.

Since the human Doppler information is modulated in the phase of a complex exponential wave (\ref{psi_l}), (\ref{2D_discrete_beat_exp}), once the correct vectors are located, a phase extraction process is performed to estimate $\{\psi_b\left[l\right]\}_{l=1}^L$ of the corresponding human (Fig. \ref{block_diagram}, block $2$.e). Traditionally, phase extraction is carried out by the arctangent demodulation algorithm \cite{park2007arctangent}, followed by an unwrapping procedure. We note that to compensate for the bounds restriction of the arctangent function, \cite{antolinos2020cardiopulmonary} and \cite{wang2020remote} used the extended differentiate and cross-multiply (DACM) algorithm \cite{wang2013noncontact} which is based on the derivative of the arctangent function. However, it is highly sensitive to noise and computationally heavy. To improve the extraction quality due to the non-linearity of the operators, \cite{sacco2020fmcw} and \cite{alizadeh2019remote} performed correction algorithms for DC component offset, prior to this step (Fig. \ref{block_diagram}, block $2$.d). Though, as we discuss in Subsection \ref{Proposed_algorithm}, this operation is not required with proper processing.

Given the extracted phase terms, where each contains an estimate of human thoracic vibration $\{v\left[ l \right]\}_{l=1}^L$ via (\ref{psi_l}), frequency estimation techniques are used to find the desired rates of respiration and heartbeat. The most common method is to apply the FFT algorithm to the given samples, relying on the fact that, at rest, the frequency bands of heartbeat and respiration are usually distinct \cite{mercuri2019vital,sacco2020fmcw,adib2015smart,alizadeh2019remote,antolinos2020cardiopulmonary}. Particularly, \cite{adib2015smart}, \cite{alizadeh2019remote} and \cite{antolinos2020cardiopulmonary} used band-pass filters to separate the domains, and enhance the relatively weak heartbeat signal. 

Aiming to improve frequency resolution and to avoid spectral leakage, 
\cite{turppa2020vital} performed zero-padding prior to the \textit{slow-time} FFT. Although extremely fine FFT resolution can be obtained this way, two coarsely separated frequencies cannot be resolved. In contrast, Adib {\it{et al.}} \cite{adib2015smart} performed linear regression on the phase of a filtered complex time-domain signal, to obtain more precise measurements. The latter improves the accuracy of the estimates, however, the maximal peak is limited by the underlying frequency grid of the DFT. Finally, \cite{kim2018low} used the MUSIC algorithm, to estimate the vital frequencies. However, this approach is sensitive to a low signal-to-noise-ratio (SNR) and suffers from high computational complexity since it requires explicit eigenvalue decomposition of an autocorrelation matrix, followed by a linear search over a large space \cite{arbenz2012lecture,das2018fast}.

\section{NCVSM of Multiple People: Extended Model and Sparsity Based Solution}
\label{Proposed}
In this work, we examine simultaneous vital signs monitoring of several people located at different radial distances from the radar, in a realistic environment that contains clutter and noise. To this end, we develop an extension to the model in (\ref{2D_discrete_beat_exp}) for FMCW radar signals addressing NCVSM of multiple people, using only a single channel and a SISO configuration. We then detail our proposed sparsity based methodology using this model. 

\subsection{Extended FMCW Model for NCVSM of Multiple People}
\label{signal_model}
Observing the model in (\ref{2D_discrete_beat_exp}), one sees that a monitored individual is characterized by frequency and phase terms of a complex wave, that respectively correspond to his distance from the radar (\ref{beat_freq}) and his thoracic motion pattern (\ref{psi_l}), for $L$ given frames. In the case of multiple people and clutter, each object in the radar's FOV, whether vibrating or stationary, can be modelled based on (\ref{2D_discrete_beat_exp}) using an appropriate beat frequency and phase. Consequently, the measured FMCW signal in this case consists of a combination of several signal reflections. To this end, we extend the signal model in (\ref{2D_discrete_beat_exp}) to
\begin{equation}
\label{y_exp_gen}
{y\left[ {n,l} \right] = \sum\limits_{m = 1}^M {{x_m}\exp \left( {j\left( {2\pi {f_m}n{T_f} + {\psi _m}\left[ l \right]} \right)} \right)}}+ w\left[ {n,l} \right],
\end{equation}
where $n=1,\ldots,N$, $l=1,\ldots,L$ and $ \{w[n,l]\}$ is a 2-D sequence of zero mean white complex Gaussian noise with some variance ${\sigma}^2$. Therefore, the received 2-D beat signal $y\left[ {n,l} \right]$ is composed of $M \leq N$ components where the triplets $\{ {x_m},{f_m},{\psi _m}\left[ l \right]\} $ denote the amplitude, frequency and phase of each $m$'th complex wave. 

The frequency $f_m$ of each object is proportional to its radial distance from the radar, $d_m$, by
\begin{equation}
    \label{fm_dm}
{f_m} \triangleq \frac{{2S}}{c}{d_m},\quad m=1,\ldots,M.
\end{equation}
We note that each frequency is distinct which complies with the SISO limitation that allows for only a single object detection at any radial distance from the radar. The \textit{slow-time} varying phase ${\psi _m}\left[ l \right]$ of each component is given by
\begin{equation}
    \label{psi_l_human}
\psi _m\left[ l \right] \triangleq \frac{{4\pi }}{{{\lambda _{\max }}}}\left( {d_m + v_m\left[ l \right]} \right),\quad l=1,\ldots,L,
\end{equation}
where the vibration function $v_m\left[ l \right]$  is generally modeled for both human and clutter objects by
\begin{equation}
\label{vib_func}
v_m\left[ l \right] \triangleq \sum\limits_{q = 1}^Q {a_{m,q}\cos \left( {2\pi g_{m,q}l{T_s}} \right)},\quad l=1,\ldots,L.
\end{equation}
The pairs $\{a_{m,q},g_{m,q}\}_{q=1}^Q$ are the corresponding amplitudes and frequencies, with the latter being limited by the \textit{slow-time} frame rate $f_s\triangleq 1/T_s$ according to $\{ g_{m,q} \}_{q=1}^Q \in \left[0\hspace{0.2cm}f_s/2\right)$, for $m=1,...,M$. 

We note that while most works did not model the vibration function, \cite{mercuri2019vital} and \cite{kim2018low} used a sum of $2$ sines corresponding to the rates of respiration and heartbeat, at a given time window. Here, the vibration model in (\ref{vib_func}) is based on generic $\{a_{m,q}\}_{q=1}^Q$ and $\{g_{m,q}\}_{q=1}^Q$ sets, that allow for adequate representation of both static and vibrating objects, such as fans. Particularly, for the case of $K$ people located in the radar's FOV, we assume that the frequency set $\{g_{m,q}\}_{q=1}^Q$ includes their HR and RR, which are denoted by $f_h^{(k)}$ and $f_r^{(k)}$, respectively, for $k=1,...,K$. As shown in Subsection \ref{Supp_rec}, the objects can be distinguished by utilizing sparse properties of the input signal, within the frequency limits that characterize normal respiration and heartbeat. 

Define the \textit{slow-time} varying complex amplitude  ${\tilde x_m}\left[ l \right]$, for $m=1,...,M$, as
\begin{equation}
    \label{x_m_l}
{\tilde x_m}\left[ l \right]\triangleq {x_m}\exp \left( {j{\psi _m}\left[ l \right]} \right),\quad l = 1,...,L.
\end{equation}
Then, (\ref{y_exp_gen}) can be represented as
\begin{equation}
    \label{y_exp_gen_2}
y\left[ {n,l} \right] = \sum\limits_{m = 1}^M {{{\tilde x}_m}\left[ l \right]\exp \left( {j2\pi {f_m}n{T_f}} \right)} +  w\left[ {n,l} \right].
\end{equation}
For each $l=1,...,L$, we can assemble the \textit{fast-time} samples of $y\left[ {n,l} \right]$ into a vector, resulting in
\begin{equation}
\label{yl_Axl}
{{{\bf{y}}_l} = {\bf{A}}{{{\bf{\tilde x}}}_l} + {{\bf{w}}_l},\quad l = 1,\ldots,L},
\end{equation}
where ${{\bf{y}}_l} \triangleq {\left[ {y\left[ {1,l} \right],...,y\left[ {N,l} \right]} \right]^T}\in \mathbb{C}^{N}$, ${\bf{A}}\in \mathbb{C}^{N\times M}$ is a Vandermonde matrix, whose entries are given by ${\bf{A}}\left( {n,m} \right) \triangleq \exp \left( {j2\pi {f_m}n{T_f}} \right)$, ${{\bf{\tilde x}}_l} \triangleq {\left[ {{{\tilde x}_1}\left[ l \right],...,{{\tilde x}_M}\left[ l \right]} \right]^T}\in \mathbb{C}^{M}$ and ${{\bf{w}}_l} \triangleq {\left[ {w\left[ {1,l} \right],...,w\left[ {N,l} \right]} \right]^T}\in \mathbb{C}^{N}$ is the noise vector.

In order to perform continuous NCVSM, the FMCW radar should operate and generate data frames throughout the entire monitoring duration. To this end, at each predefined time interval $T_{\textrm{int}}$, the sequence $\{{\bf{y}}_l\}_{l=1}^L$ (\ref{yl_Axl}) is formed by collecting the last $L$ frames up to that point in time. The number of frames to be processed, $L$, is determined by a predefined time window ${T_{\textrm{win}}}$ according to $L={T_{\textrm{win}}}f_s$, where the units of ${T_{\textrm{win}}}$ and $f_s$ are $[s]$ and $[1/s]$, respectively. For convenience, we reformulate the observations in (\ref{yl_Axl}) for each ${T_{\textrm{int}}}$, using the following matrix form
\begin{equation}
    \label{Y_mat=AX_mat+W_mat}
{\bf{Y}}= {\bf{A}}{\bf{\tilde{X}}}+{\bf{W}},
\end{equation}
where ${\bf{Y}}\triangleq\left[ {{{\bf{y}}_1},...,{{\bf{y}}_L}} \right]\in \mathbb{C}^{N\times L}$ is the given measurement matrix, ${\bf{\tilde{X}}} \triangleq \left[ {{{{\bf{\tilde x}}}_1},...,{{{\bf{\tilde x}}}_L}} \right]\in \mathbb{C}^{M\times L}$ and ${\bf{W}} \triangleq \left[ {{{\bf{w}}_1},...,{{\bf{w}}_L}} \right]\in \mathbb{C}^{N\times L}$ is the noise matrix. The above data model is assumed to have the following properties:
\begin{enumerate}[{A}-1]
    \item  The sequence  $\left\{ {{{{\bf w}}_l}} \right\}_{l = 1}^L$ which forms the noise matrix ${\bf{W}}$, can be viewed as $L$ i.i.d realizations of a zero mean complex Gaussian noise vector with covariance matrix ${\sigma ^2}{\bf{I}}_N$, where ${\bf{I}}_N$ denotes a size-$N$ identity matrix.
    \item 
    Only $K\ll M$ humans are being monitored, where each is stationary except for minor movements caused by breathing or speaking. The latter induces a row-wise sparsity in ${\bf{\tilde{X}}}$, meaning that the vectors $\{{\bf{\tilde{x}}}_l\}_{l=1}^L$ share a joint support.
\end{enumerate}

Based on the model in (\ref{Y_mat=AX_mat+W_mat}), the first goal is to recover the row-coordinates of ${\bf{\tilde{X}}}$ associated with humans in the radar's FOV, denoted by the support ${{\cal S}}$, by which we can estimate the human locations and filter the relevant Doppler information. The second goal, using the recovered ${{\cal S}}$, is to continuously evaluate the RR and HR of each detected human throughout the complete monitoring duration. Mathematically, every $T_{\textrm{int}}$ seconds we seek to estimate the corresponding $f_r^{(k)}$ and $f_h^{(k)}$, for $k=1,...,K$.

\subsection{Sparsity Based NCVSM of Multiple People}
\label{Proposed_algorithm}
With the aid of the upper diagram of Fig. \ref{block_diagram}, below we detail each stage of the proposed sparsity based approach, based on the model developed in Subsection \ref{signal_model}.

 \subsubsection{Pre Processing}
\label{Pre_Proc}
At the start of each monitoring iteration determined by $T_{\textrm{int}}$, we perform a preliminary processing to assemble ${\bf{Y}}$ according to (\ref{Y_mat=AX_mat+W_mat})  with an increased SNR (Fig. \ref{block_diagram}, block $1$.a).

Recall from Assumption A-1, that each element of ${\bf{W}}$ is derived from a Gaussian distribution with variance $\sigma ^2$. Hence, we can utilize the slowness of thoracic motion relative to a single chirp \cite{alizadeh2019remote} to reduce the variance of each element, by averaging several data observations at each frame. To accomplish this, we define a transmission scheme in which $G>1$ consecutive chirps are transmitted in each frame instead of a single one, with $G$ being limited by the frame duration $T_s$, as illustrated in Fig. \ref{fig:saw-tooth}.

For every $T_{\textrm{int}}$, this process generates $G$ beat signal duplicates per frame that differ only in the noise impact, i.e., arranging the input data similalry to (\ref{Y_mat=AX_mat+W_mat}) results in ${\bf{Y}}_{\textrm{pre}}=\left[ \{{{{\bf{y}}_{1,g}}\}_{g=1}^G,...,\{{{\bf{y}}_{L,g}}}\}_{g=1}^G \right]\in \mathbb{C}^{N\times GL}$. By averaging the \textit{fast-time} row samples every $G$ columns, we get ${\bf{Y}}=\left[ {{\bar{{\bf{y}}}_1},...,{\bar{{\bf{y}}}_L}} \right]\in \mathbb{C}^{N\times L}$, where ${\bar{{\bf{y}}}_l} \triangleq \frac{1}{G}\sum\limits_{g = 1}^G {{\bf{y}}_{l,g}}$, $l=1,...,L$. This procedure yields the model in (\ref{Y_mat=AX_mat+W_mat}) while reducing the noise variance of each data element by a factor of $G$, v. to the use of a single chirp per frame.
\subsubsection{Support Recovery and Human Localization}
\label{Supp_rec}
Here, we estimate the support of ${\bf{\tilde{X}}}$, denoted $\cal S$, which allows to detect the radial distance of each human from the radar by (\ref{fm_dm}), and to efficiently extract the corresponding Doppler samples for the remainder of the monitoring process, as detailed in Subsection \ref{Dopp_Rows_Rec}. Additionally, we show that this localization can be performed only once in the entire monitoring period (Fig. \ref{block_diagram}, block 1.b).

To this end, first, the $\textit{fast-time}$ frequencies $\{f_m\}_{m=1}^M$
(\ref{fm_dm}) are assumed to lie on the Nyquist grid, i.e.,
\begin{equation}
    \label{fm_Nyquist_grid}
f_m  =\frac{{{f_{\textrm{ADC}}}}}{N} {i_m}{\rm{,}}\quad {\rm{ }}{i_m} = 0,...,M - 1,
\end{equation}
where $f_{\textrm{ADC}}\triangleq1/T_f$ is determined by the ADC component. We note using (\ref{fm_dm}) and (\ref{fm_Nyquist_grid}), that the maximal detectable distance is $d_{\textrm{max}}=\frac{cf_{\textrm{ADC}}}{{2SN}}\left(M-1\right)$. Interestingly, since $N$ is the number of \textit{fast-time} samples, generally, $M = N$, although as detailed in Subsection \ref{Dopp_Rows_Rec}, by selecting $M = N/2$ we can employ the model in (\ref{Y_mat=AX_mat+W_mat}) even using data from only a single channel, without jeopardizing estimation performance.

Second, we denote by $B^{(R)}$ and $B^{(H)}$ the frequency bands of normal respiration and heartbeat, respectively, where
${B^{\left( R \right)}},{B^{\left( H \right)}} \in \left[0\hspace{0.2cm}f_s/2 \right)$. This prior knowledge of human-typical pulse and breathing frequencies, aids in the separation of humans from static or vibrating clutter, such as fans. Hence, based on the \textit{slow-time} frequency modulation structure of the vibration signal $v_m\left[l\right]$ (\ref{psi_l_human}), (\ref{vib_func}) in (\ref{y_exp_gen}), we perform spectral filtering of $\bf{Y}$ in the \textit{slow-time} axis, according to 
\begin{equation}
    \label{Y_filtered}
{\bf{\bar Y}} = \frac{1}{L}{\left( {{\bf{F}}_L^H\left( {{\bf{\Pi }} \odot {{\bf{F}}_L}{{\bf{Y}}^T}} \right)} \right)^T},
\end{equation}
where ${{\bf{F}}_L}$ is a full $L$-size DFT matrix, ${\bf{\Pi }}$ denotes an ideal window corresponding to the vital frequencies in ${B^{\left( H \right)}} \cup {B^{\left( R \right)}}$ and $\odot$ denotes the element-wise product.

Since by Assumption A-2, ${\bf{\tilde{X}}}$ is a row-sparse matrix, inspired by \cite{rossman2019rapid}, we now recover it from ${\bf{\bar Y}}$ using a JSR technique formulated by the following optimization problem 
\begin{equation}
    \label{X_rec_JSR}
\mathop {\min }\limits_{{\bf{\tilde X}} \in \mathbb{C}{^{M \times L}}} \;\left\| {{\bf{\bar Y}} - {\bf{A\tilde X}}} \right\|_F^2 + \lambda {\left\| {{\bf{\tilde X}}} \right\|_{2,1}}.
\end{equation}
Here, to promote the row sparsity of ${\tilde{\bf{X}}}$, we use the regularization parameter $\lambda\geq 0$ and the mixed $l_{2,1}$ norm defined by $\|{\bf{{X}}}\|_{2,1}\triangleq \sum _i \|{\bf{x}}^i\|_2$, with ${\bf{x}}^i$ denoting the $i$'th row of a matrix ${{\bf{X}}}$. Similarly to \cite{rossman2019rapid}, we solve (\ref{X_rec_JSR}) and find the support $\cal{S}$ using the fast iterative soft-thresholding algorithm (FISTA) \cite{beck2009fast,palomar2010convex}.

The recovered support ${\cal S}$ is first used to calculate the distances of the monitored people from the radar through (\ref{fm_dm}), that is, $\{d_m\}$, for $ m \in {\cal S}$. Then, since we are studying the case of monitoring stationary subjects, the coordinates of ${\cal S}$ are fixed throughout the monitoring, implying that we can recover ${\cal S}$ only once, and use it for all subsequent iterations.

\subsubsection{Doppler Rows Recovery}
\label{Dopp_Rows_Rec}
The support evaluated in the previous step allows us to efficiently recover only the human-related Doppler samples of ${\bf{\tilde{X}}}$ given ${\bf{Y}}$, throughout the remainder of the monitoring process (Fig. \ref{block_diagram}, block 1.c).

Using $\cal{S}$ and Assumption A-2, the model in (\ref{Y_mat=AX_mat+W_mat}) can be written as
\begin{equation}
    \label{Y=AsXs+W}
    {\bf{Y}}= {\bf{A}}_{\cal{S}}{\bf{\tilde{X}}}_{\cal{S}}+{\bf{W}},
\end{equation}
with ${{\bf{A}}_{{\cal S}}} \in \mathbb{C}{^{N \times{K}}}$ and ${{\bf{\tilde{X}}}_{{\cal S}}} \in \mathbb{C}{^{K\times L}}$ respectively being the atoms of $\bf{A}$ and the rows of ${\bf{\tilde{X}}}$ corresponding to $\cal{S}$, where $K= \left|\cal{S}\right|$. By knowing the support for each $T_{\textrm{int}}$, we can directly estimate ${{\bf{\tilde{X}}}_{{\cal S}}}$ from ${\bf{Y}}$, using the solution of the following Least-Squares (LS) problem \cite{ruppert1994multivariate}
\begin{equation}
    \label{F-LS}
{ \mathop {\min }\limits_{{\bf{\tilde{X}}}_{{\cal S}}\in \mathbb{C}{^{K \times L}}} \;\left\| {{\bf{ Y}}- {{\bf{A}}_{{\cal S}}}{{{\bf{\tilde{X}}}}_{{\cal S}}}} \right\|_F^2},
\end{equation}
given by
\begin{equation}
    \label{X_s_estimate}
{{{{{\widehat{{\bf{X}}}}_{{\cal S}}}}} = {{\left( {{\bf{A}}_{{\cal S}}^H{{\bf{A}}_{{\cal S}}}} \right)}^{ - 1}}{\bf{A}}_{{\cal S}}^H{\bf{Y}}}.
\end{equation}
By the Vandermonde structure of $\bf{A}$ defined below (\ref{yl_Axl}) and since $K\ll M \leq N$, we have that ${\bf{A}}_{{\cal S}}^H{{\bf{A}}_{{\cal S}}}$ is invertible. Explicitly, ${\bf{A}}_{{\cal S}}={\bf{F}}_{{\cal S}}^H$ and ${\bf{A}}_{{\cal S}}^H{{\bf{A}}_{{\cal S}}}=N{\bf{I}}_K$, meaning that ${{{{\widehat{{\bf{X}}}}_{{\cal S}}}}}$ in (\ref{X_s_estimate}) can be represented as
\begin{equation}
    \label{X_s_estimate_final}
{{{{{\widehat{{\bf{X}}}}_{{\cal S}}}}} = \frac{1}{N}{\bf{F}}_{{\cal S}}{\bf{ Y}}},
\end{equation}
where ${\bf{F}}_{{\cal S}}$ denotes a partial DFT matrix corresponding to the \textit{fast-time} frequencies $\{f_m\}$ (\ref{fm_Nyquist_grid}), for $m \in \cal{S}$.

Note that by assuming $M=N/2$, the frequencies $\{f_m\}_{m=1}^M$ (\ref{fm_Nyquist_grid}) correspond to the positive tones of the sinusoidal combination $\sum\limits_{m = 1}^M {{x_m}\cos \left( { {2\pi {f_m}n{T_f} + {\psi _m}\left[ l \right]} } \right)}$, which would have been obtained in (\ref{y_exp_gen}) when using the \textit{In-Phase} channel solely. Hence, for $M=N/2$ the estimator in (\ref{X_s_estimate_final}) when using both the I and Q channels is equivalent to that obtained from the use of only a single channel (I or Q), up to a constant factor. To avoid hardware overload and potential issues of using both channels, we assume here that $M=N/2$ and use only the \textit{In-Phase} channel.

As opposed to existing methods that compute the entire \textit{Range vs. Slow-Time} map (Fig. \ref{block_diagram}, block $2$.b), which is equivalent to replacing ${\bf{F}}_{\cal{S}}$ in (\ref{X_s_estimate_final}) with ${\bf{F}}_M$ that corresponds to all $M=N/2$ frequencies in (\ref{fm_Nyquist_grid}), at each iteration the proposed estimator recovers only the relevant DFT samples, which is considerably more efficient since $\left|{\cal{S}}\right|=K\ll M$. 

Finally, Since there are no humans within $d_m=0$ meters of the radar, the estimator in (\ref{X_s_estimate_final}) is always filtering the DC component, corresponding to $m=1$ in (\ref{fm_Nyquist_grid}). As a result, we do not include a mechanism for DC offset correction unlike other techniques (Fig \ref{block_diagram}, block 2.d).
\subsubsection{Phase Extraction}
\label{phase_extr}
Using relation (\ref{x_m_l}) and the definition of ${\bf{\tilde{X}}}$ below (\ref{Y_mat=AX_mat+W_mat}), we have that
\begin{equation}
    \label{X_s}
    {\bf{\tilde{X}}}_{\cal{S}}\left(m,l\right)={x_m}\exp \left( {j{\psi _m}\left[ l \right]} \right),\quad m \in {\cal S}.
\end{equation}
That is, $\widehat{{\bf{X}}}_{\cal{S}}$ (\ref{X_s_estimate_final}) estimates the \textit{slow-time} varying phasor terms associated with humans in the radar's FOV. In order to estimate the appropriate RR and HR from the thoracic vibrations of each individual, i.e., $\{f_r^{(k)},f_h^{(k)}\}_{k=1}^K$ from $\{v_m[l]\}_{l=1}^L$, $m \in {\cal S}$, we must first extract an approximation of the phase terms $\{\psi_m\left[ l \right]\}_{l=1}^L$, $m \in \cal{S}$ (\ref{psi_l_human}) from $\widehat{{\bf{X}}}_{\cal{S}}$ (Fig. \ref{block_diagram}, block $1$.d). Hence, we perform the following element-wise angle extraction operation on ${{{{\widehat{{\bf{X}}}}_{{\cal S}}}}}$, which yields the following $L\times{K}$ vibration matrix
\begin{equation}
    \label{V_l_k}
{\bf{V}}\left(l,k\right) \triangleq unwrap\left( {\angle \left( {{{\widehat{{\bf{X}}}}_{S}}}\left({{\cal S}\{ k}\},l\right)  \right)} \right)^{T},\hspace{0.2cm} {\begin{cases} k=1,...,K \\l=1,...,L \end{cases}}
\end{equation}
where $unwrap\left(\cdot\right)$ denotes the unwrapping procedure based on \cite{alizadeh2019remote}, used since the unambiguous phase range is limited by $(-\pi\hspace{0.1cm}\pi]$. 
The angle extraction operator $\angle\left(\cdot\right)$ is based on the four quadrant arctangent function applied with Matlab's "atan2.m" function.

\subsubsection{VSDR Method for Estimating RR and HR}
\label{RR_HR_estimation}
In the final stage of each iteration, both the RR and HR of each individual, $\{f_r^{(k)},f_h^{(k)}\}_{k=1}^K$, are estimated given the vibration matrix $\bf{V}$, and recorded at the corresponding time instant defined by $T_{\textrm{int}}$ (Fig. \ref{block_diagram}, block $1$.e). Below, we develop a procedure for selecting high-resolution estimates of vital signs out of two unique dictionaries based on human-typical pulse and breathing frequencies.

First, using (\ref{psi_l_human}) and (\ref{vib_func}), the matrix $\bf{V}$ in (\ref{V_l_k}) can be viewed as a chain of vectors corresponding to the detected humans' thoracic vibration signals, i.e.,
\begin{equation}
    \label{V_final}
{\bf{V}} = \left[ { {\bf{v}}_1, \ldots ,{\bf{v}}_K } \right],
\end{equation}
with
\begin{equation}
    \label{v_k}
{\bf{v}}_k= {\bf{D}}{{\bf{a}}_k}+{{\bf{n}}_k},\quad k=1,\ldots,K,
\end{equation}
where each vector ${{\bf{a}}_k}\in\mathbb{R}{^Q}$ consists of $Q$ amplitudes $\{{\tilde a}_q^{(k)}\}_{q=1}^Q$ and ${\bf{D}} \in \mathbb{R}{^{L \times Q}}$ is a cosine-based dictionary matrix with entries ${\bf{D}}\left( {l,q} \right) \triangleq \cos \left( {2\pi \tilde{g}_q l{T_s}} \right)$ where the frequencies $\{\tilde{g}_q\}_{q=1}^Q \in \left[0\hspace{0.2cm}f_s/2\right) $. Finally, $\{{\bf{n}}_k\}_{k=1}^K$ denotes a sequence of length-$L$ i.i.d. noise vectors as a result of the non-linear operations in (\ref{V_l_k}).

Ideally, to achieve optimal frequency resolution, one has to use $T_{\textrm{win}}=60$ seconds which correspond to the number of heartbeats or breath cycles per minute definition, i.e., bpm. However, this comes at the expense of a reduced temporal localization. Therefore, to allow for smaller time windows but with increased resolution we uniformly divide the segment $\left[0\hspace{0.2cm}f_s/2\right)$ according to a resolution of $1$ bpm, i.e., the frequencies $\{\tilde{g}_q\}_{q=1}^Q$ satisfy
\begin{equation}
    \label{g_q_freqs}
\tilde{g}_q = {h_q}\frac{{{f_s}}}{Q}{\rm{,}}\quad {\rm{ }}{h_q} = 0,...,\frac{Q}{2} - 1,\quad Q=60f_s.
\end{equation}

We assume that for every $T_{\textrm{win}}$, the $2$ most dominant frequencies of each thoracic vibration ${\bf{v}}_k$, are the rates of heartbeat and respiration. However, the amplitude of the heart signal is much smaller than that of respiration, so in order to facilitate its detection, similarly to \cite{mercuri2019vital,sacco2020fmcw,adib2015smart,alizadeh2019remote,antolinos2020cardiopulmonary}, we utilize the phenomenon that at rest, the frequency bands are usually separated from each other. We note that unlike \cite{adib2015smart,alizadeh2019remote} and \cite{antolinos2020cardiopulmonary}, we do not perform a signal separation procedure prior to this stage (Fig. \ref{block_diagram}, block $2$.f). Here, we exploit this band-separation property to effectively focus the frequency search on limited dictionaries corresponding to each band. Explicitly, we define the vital signs based dictionaries ${\bf{D}}^{\left( R \right)}\in \mathbb{R}{^{L \times Q_R}}$ and ${\bf{D}}^{\left( H \right)}\in \mathbb{R}{^{L \times Q_H}}$ as follows
\begin{equation}
    \label{VSDR_def}
\begin{array}{l}
{{\bf{D}}^{\left( R \right)}}\left( {l,q} \right) \triangleq \cos \left( {2\pi \tilde g_q^{\left( R \right)}l{T_s}} \right)\hspace{0.2cm}\textrm{and}\\
{{\bf{D}}^{\left( H \right)}}\left( {l,q} \right)\triangleq \cos \left( {2\pi \tilde g_q^{\left( H \right)}l{T_s}} \right),
\end{array}
\end{equation}
where both the frequencies $\{\tilde{g}_q^{\left( R \right)}\} $ and $\{\tilde{g}_q^{\left( H \right)}\} $ constitute a subset of (\ref{g_q_freqs}), satisfying $\{\tilde{g}_q^{\left( R \right)}\} \in B^{\left( R \right)}$ and $\{\tilde{g}_q^{\left( H \right)}\} \in B^{\left( H \right)}$. 

Using the notation from (\ref{VSDR_def}) and the assumption that only the vital frequency bands $B^{\left( R \right)}$ and $B^{\left( H \right)}$ contribute to the thoracic vibration of each ${\bf{v}}_k$, we can represent the model in (\ref{v_k}) as
\begin{equation}
    \label{v_k_VSDR}
{{\bf{v}}_k} = {{\bf{D}}^{\left( R \right)}}{\bf{a}}_k^{\left( R \right)} + {{\bf{D}}^{\left( H \right)}}{\bf{a}}_k^{\left( H \right)} + {{\bf{n}}_k},\quad k=1,\ldots,K,
\end{equation}
where ${\bf{a}}_k^{\left( R \right)}\in\mathbb{R}{^{Q_R}}$ and ${\bf{a}}_k^{\left( H \right)}\in\mathbb{R}{^{Q_H}}$ are both amplitude vectors with only a single non-zero element whose coordinate indicates the corresponding rate (RR/HR), of the $k$-th human.

To recover ${\bf{a}}_k^{\left( R \right)}$ and ${\bf{a}}_k^{\left( H \right)}$ from each ${{\bf{v}}_k}$ in (\ref{v_k_VSDR}), we apply the following estimators
\begin{equation}
    \label{VSDR_a_k_rec}
{\bf{\hat a}}_k^{\left( R \right)} = {{\bf{D}}^{\left( R \right)T }}{{\bf{v}}_k}\hspace{0.2cm}\textrm{and}\hspace{0.2cm}{\bf{\hat a}}_k^{\left( H \right)} = {{\bf{D}}^{\left( H \right)T }}{{\bf{v}}_k},
\end{equation}
$\forall k=1,...,K$. The maximum's coordinate of ${\bf{\hat a}}_k^{\left( R \right)}$ and ${\bf{\hat a}}_k^{\left( H \right)}$ points to the RR estimation $\hat{f}_r^{(k)}$ and the HR estimation $\hat{f}_h^{(k)}$, from $\{\tilde{g}_q^{\left( R \right)}\}$ and $\{\tilde{g}_q^{\left( H \right)}\}$, respectively. To enhance estimation stability, we replace the computed $\hat{f}_r^{(k)}$ and $\hat{f}_h^{(k)}$ with the average of the estimates from the last $3$ and $1.5$ seconds, respectively. This vital signs based dictionary recovery is referred to as the VSDR method.

Algorithm 1 summarizes our approach with $L_{\textrm{lip}}$ and $I_{\textrm{max}}$ respectively denote the Lipschitz constant, and the maximal number of iterations in the FISTA algorithm \cite{beck2009fast}, and the set of all measurement matrices (\ref{Y_mat=AX_mat+W_mat}) being processed during the monitoring is denoted by $\{\bf{Y}\}$. 

\begin{algorithm}[h!]
\label{alg}
\begin{algorithmic}
\caption{Sparsity Based NCVSM of Multiple People}
\State \hspace{-0.4cm} \textbf{Input:} $T_{\textrm{int}}$, $\{{\bf{Y}}\}$, $\bf{A}$, $\lambda\geq 0$, $L_{\textrm{lip}}$, $I_{\textrm{max}}$, ${\bf{D}}^{\left( R \right)}$, ${\bf{D}}^{\left( H \right)}$.
\State \hspace{-0.3cm}\textbf{At each} $T_{\textrm{int}}$ \textbf{do:}
\State \textbf{First iteration:}
\State \hspace{0.3cm} \textbf{1:} Pre-process ${\bf{Y}}$ according to Subsection \ref{Pre_Proc}
\State \hspace{0.3cm} \textbf{2:} Filter spatial interference by (\ref{Y_filtered}) to obtain ${\bf{\bar Y}}$ 
\State \hspace{0.3cm} \textbf{3:} Perform JSR by (\ref{X_rec_JSR}) using FISTA \cite{beck2009fast} and save ${\cal{S}}$
\State \hspace{0.3cm} \textbf{4:} Compute the distances $\{d_m\}$, $m \in {\cal S}$ using (\ref{fm_dm})
\State \hspace{0.3cm} \textbf{5:} Recover ${{{{\widehat{{\bf{X}}}}_{{\cal S}}}}}$ given ${{\cal S}}$ and ${\bf{Y}}$ by (\ref{X_s_estimate_final})
\State \hspace{0.3cm} \textbf{6:} Extract ${\bf{V}}$ from ${{{{\widehat{{\bf{X}}}}_{{\cal S}}}}}$ according to (\ref{V_l_k})
\State \hspace{0.3cm} \textbf{7:} Estimate $\{{\bf{\hat a}}_k^{\left( R \right)},{\bf{\hat a}}_k^{\left( H \right)}\}_{k=1}^K$ given ${\bf{V}}$ by VSDR (\ref{VSDR_a_k_rec})
\Statex \textbf{Output:} $\hat{f}_r^{(k)}$ and $\hat{f}_h^{(k)}$, $\forall k=1,...,K$ 
\State {\textbf{In all other iterations:}}
\State \hspace{0.3cm} \textbf{1:} Pre-process $\bf{Y}$ and skip to steps \textbf{5-7} using ${\cal{S}}$
\Statex \textbf{Output:} $\hat{f}_r^{(k)}$ and $\hat{f}_h^{(k)}$, $\forall k=1,...,K$ 
\end{algorithmic}
\end{algorithm}

\section{Numerical Examples}
\label{Numerical}
In this section, the performance of the proposed method is evaluated and compared to existing techniques, using a simulation that combines the measurement model in (\ref{Y_mat=AX_mat+W_mat}) with real Electrocardiography (ECG) and impedance data of $30$ participants from \cite{schellenberger2020dataset}. Note that Algorithm $1$ is divided such that during the first monitoring iteration, a localization procedure is performed, after which the vital signs of the detected people are estimated throughout the rest of the monitoring period. As a result, we present here two simulation studies. The first investigates the multiple-people localization part in a clutter-rich environment, while the second examines NCVSM given human thoracic vibrations from the previous study.

To this end, we used relations (\ref{y_exp_gen})-(\ref{vib_func}) that form the model in (\ref{Y_mat=AX_mat+W_mat}), to compose seven different objects in the radar's FOV (of which $K=3$ humans), each characterized by a corresponding $x_m$ (\ref{y_exp_gen}), $d_m$ (\ref{fm_dm}) and $\{v_m\left[l\right]\}_{l=1}^L$ (\ref{vib_func}), as detailed in Table \ref{table:local_exp}. A schematic illustration of the experiment can be seen in Fig. \ref{fig:FMCW_radar_MP}. To create a realistic environment, as shown in Table \ref{table:local_exp}, we adjusted each $x_m$ so that static objects have the strongest reflections, followed by fans, and finally humans, all of which fade as a function of the distance $d_m$. Furthermore, to examine the impact of environmental noise, we used an SNR term that controls the variance of $ \{w[n,l]\}$ (\ref{y_exp_gen}) via $\textrm{SNR}\triangleq 1/{\sigma ^2}$. As to the monitoring, we determined that the humans would be monitored simultaneously for $10$ minutes, with RR and HR estimates computed every $T_{\textrm{int}}=0.05$ [s], using \textit{In-phase} channel data collected from the last $T_{\textrm{win}}=30$ [s], starting at $T_{\textrm{win}}$.

\begin{figure}[htbp!]  
\begin{center}
\subfigure[]{\label{Ref_Impedance}\includegraphics[width=0.39\textwidth]{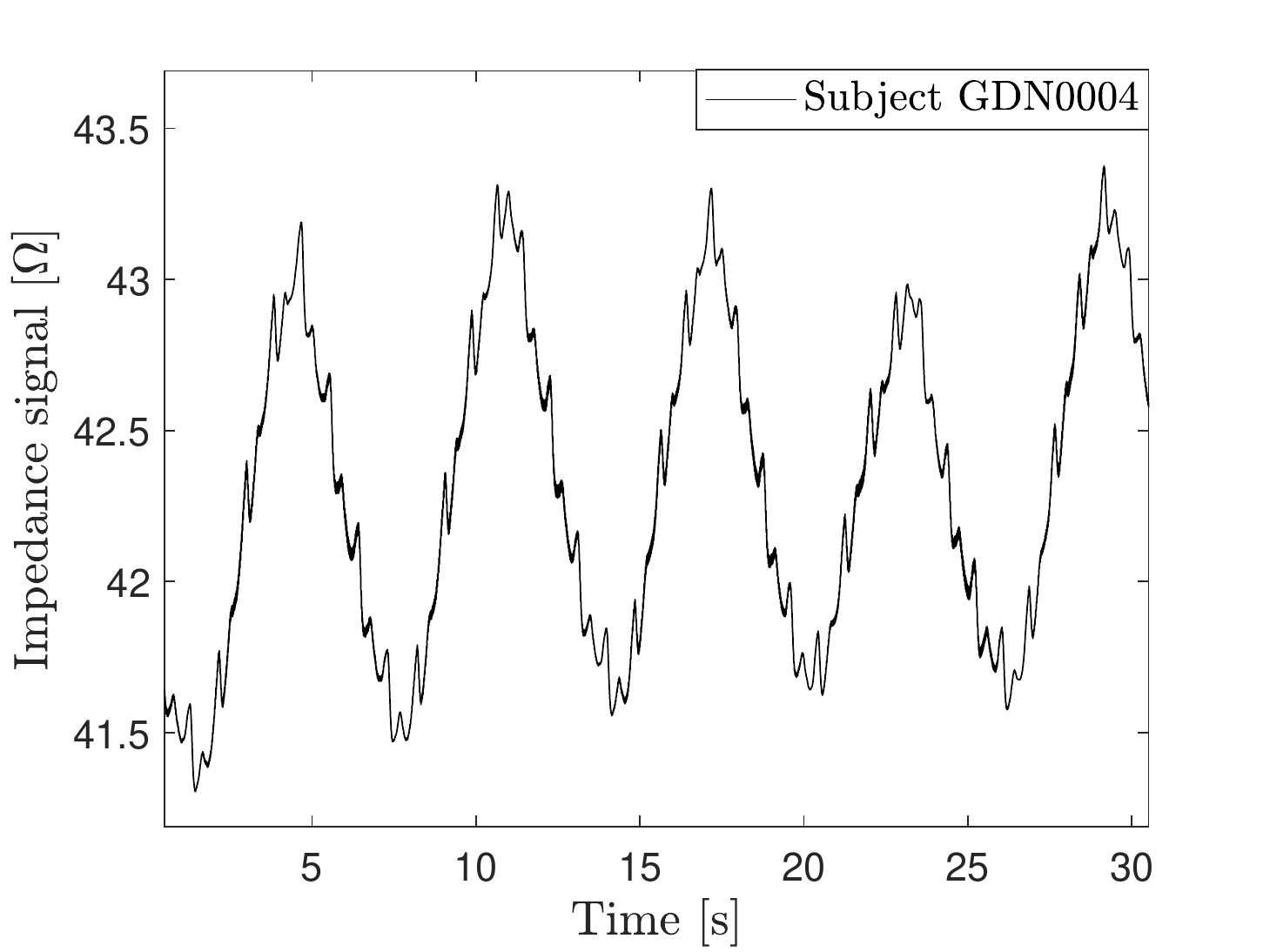}}\vspace{-0.3cm}
\subfigure[]{\label{Ref_ECG}\includegraphics[width=0.39\textwidth]{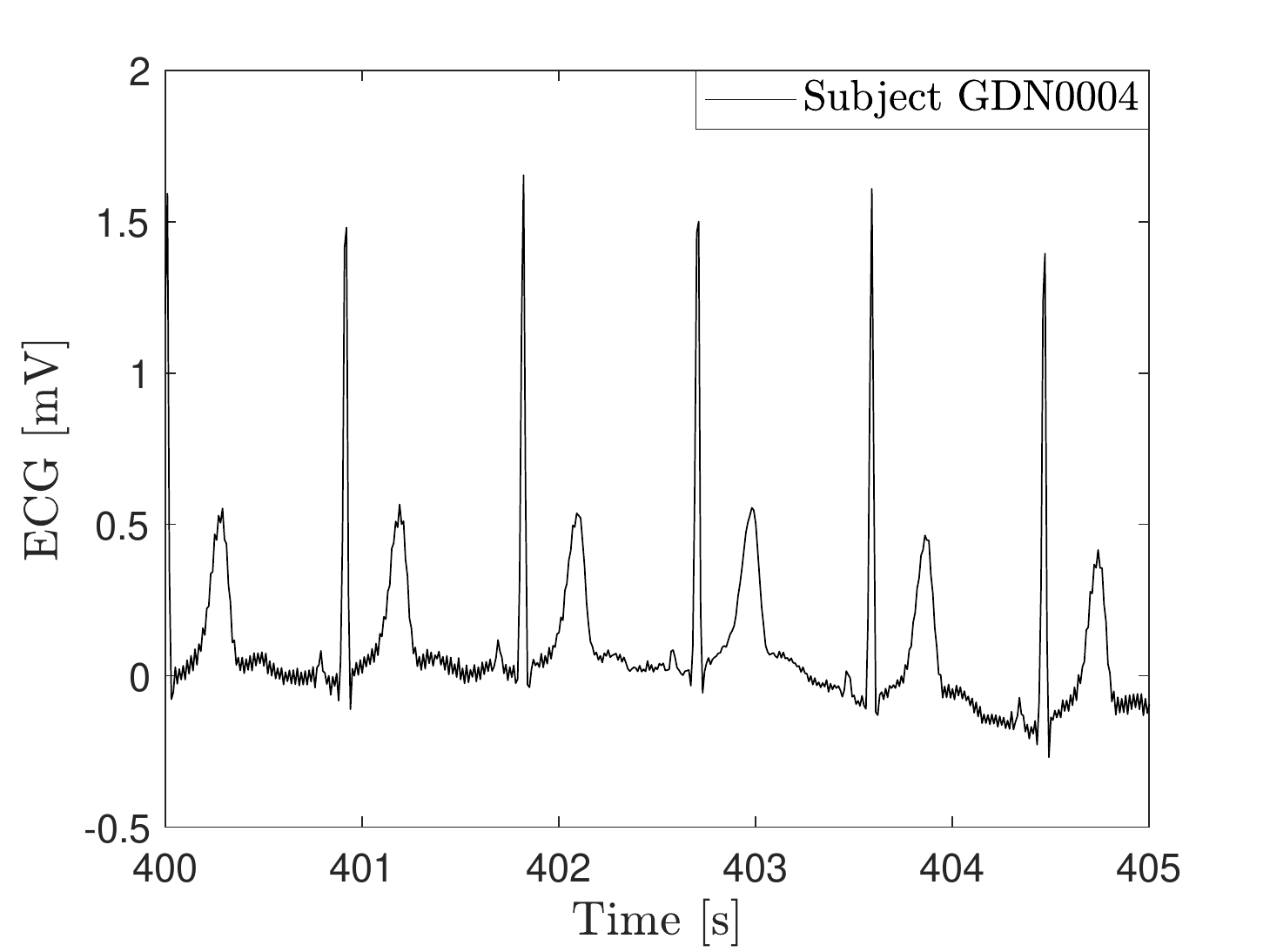}}
\end{center} 
\caption{Real Impedance and ECG data from \cite{schellenberger2020dataset}. \textbf{(a)} Impedance [$\Omega$] - used as a reference for comparing the RR estimates and to simulate human thoracic vibration. \textbf{(b)} ECG [mV] - used as a reference for comparing the HR estimates.}   
\label{fig:references}
\end{figure}

To reliably simulate prolonged human breathing, we used data from the resting scenario of \cite{schellenberger2020dataset}, in which the participants were lying on a table wired to several monitoring devices and were told to breath calm and avoid large movements for at least $10$ minutes. In our analysis, we selected and rescaled the $100$ [Hz] impedance signal "tfm$\_$z0", which provides insight into the impedance change of the thorax, to simulate human thoracic vibrations over $10$ minutes of monitoring, as shown in Fig. \ref{Ref_Impedance} and implemented in Table \ref{table:local_exp}. 
A variety of cardiac and respiratory parameters can be extracted from the impedance signal, including RR and HR \cite{ernst1999impedance, sherwood1990methodological}, thus the raw signal serves as a reference for comparing the RR estimation results. As to the HR reference, we used the gold-standard $2000$ [Hz] ECG signal "tfm$\_$ecg1" (Fig. \ref{Ref_ECG}), and down-sampled it to $100$ [Hz] to correspond to $T_s=10$ [ms]. The main FMCW radar parameters for assembling the model in (\ref{Y_mat=AX_mat+W_mat}) are based on Texas Instruments IWR1642 76 to 81 [GHz] mmWave sensor \cite{TIdatasheet}, and summarized in Table \ref{table:FMCW_parameters}.

\begin{table}[h!]
\centering
\begin{tabular}{ |l |c | c| l| }
\hline
\hspace{0.2cm} Object type & $x_m$ & $d_m$ & \hspace{1.2cm}$\{v_m\left[l\right]\}_{l=1}^L$\\
\hline
Vibrating fan $\#1$ & $0.7$   & $1.5$  & $0.1\cos{\left(2\pi40lT_s\right)}$, $l=1,...,L$ \\ 
\bf{Human $\#1$} & $0.5$  & $2$ & Impedance data from \cite{schellenberger2020dataset}\\
Static clutter $\#1$ & $1$   & $2.3$ & $0$, $\forall l=1,...,L$ \\
\bf{Human $\#2$} & $0.45$   & $2.6$ & Impedance data from \cite{schellenberger2020dataset}\\
Static clutter $\#2$ & $0.9$   & $2.9$ & $0$, $\forall l=1,...,L$ \\
Vibrating fan $\#2$ & $0.6$   & $3.1$ &  $0.1\cos{\left(2\pi40lT_s\right)}$, $l=1,...,L$ \\
\bf{Human $\#3$} & $0.4$   & $3.5$ & Impedance data from \cite{schellenberger2020dataset}\\
 \hline
\end{tabular}
\caption{Setup of multiple objects scenario}
\label{table:local_exp}
\end{table}

\begin{table}[h!]
\centering
\begin{tabular}{ |l |c | c|  }
\hline
Parameter & Symbol & Value\\
\hline
Maximal chirp wavelength & $\lambda_{\textrm{max}}$   & $3.9$ [mm]\\
Chirp duration & $T_c$   & $57$ [$\mu \textrm{s}$]\\
ADC sampling rate & $f_{\textrm{ADC}}$   & $4$ [MHz]\\
Rate of frequency sweep & $S$   & $70$ [MHz/$\mu$s]\\
Frame duration & $T_s$   & $10$ [ms]\\
$\#$ \textit{fast-time} samples  & $N$   & $200$ \\
$\#$ of chirps per frame & $G$   & $150$ \\
 \hline
\end{tabular}
\caption{FMCW radar parameters}
\label{table:FMCW_parameters}
\end{table}

\subsection{Multiple People Localization}
This study examined the localization of several people in a clutter-rich environment given the pre-processed ${\bf{Y}}$ of the first iteration, as a preliminary step to monitor their vital signs.

The parameters of the proposed JSR localization technique were set as follows. The vital \textit{slow-time} frequencies of $\bf{\Pi}$ in (\ref{Y_filtered}) were drawn from the length-$L$ Nyquist grid determined by $f_s$. The parameters for solving (\ref{X_rec_JSR}) using FISTA \cite{beck2009fast} were set to $\lambda=30$, $L_{\textrm{lip}}=4.5e6$ and $I_{\textrm{max}}=1000$. Finally, the thoracic vibrations selected here for humans in table \ref{table:local_exp}, were based on resting impedance data of subjects $1-3$ from \cite{schellenberger2020dataset}.

\begin{figure*}[htbp!]  
\begin{center}
\subfigure[]{\label{range_vs_slow_time}\includegraphics[width=0.44\textwidth]{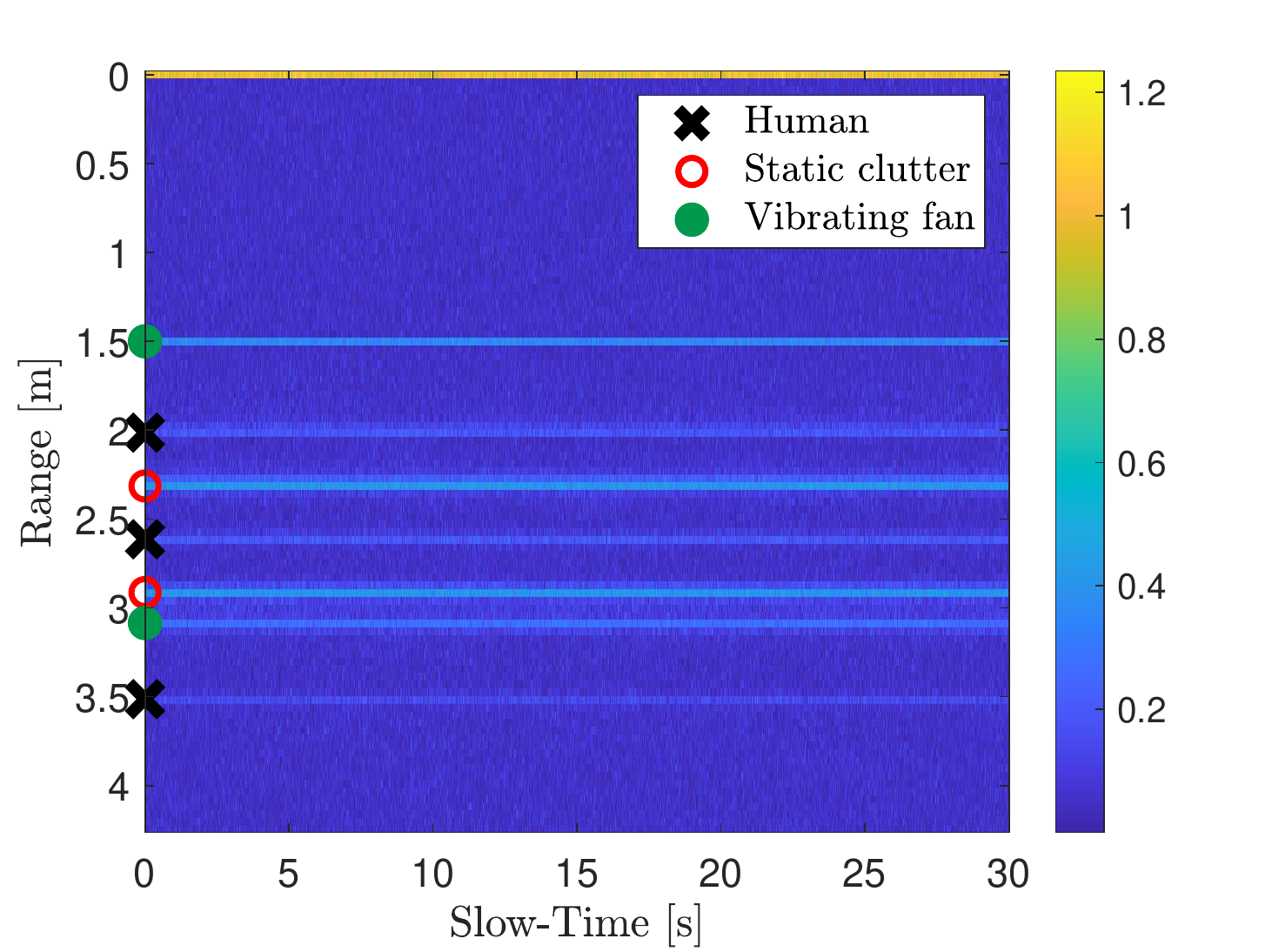}} 
\subfigure[]{\label{localization}\includegraphics[width=0.43\textwidth]{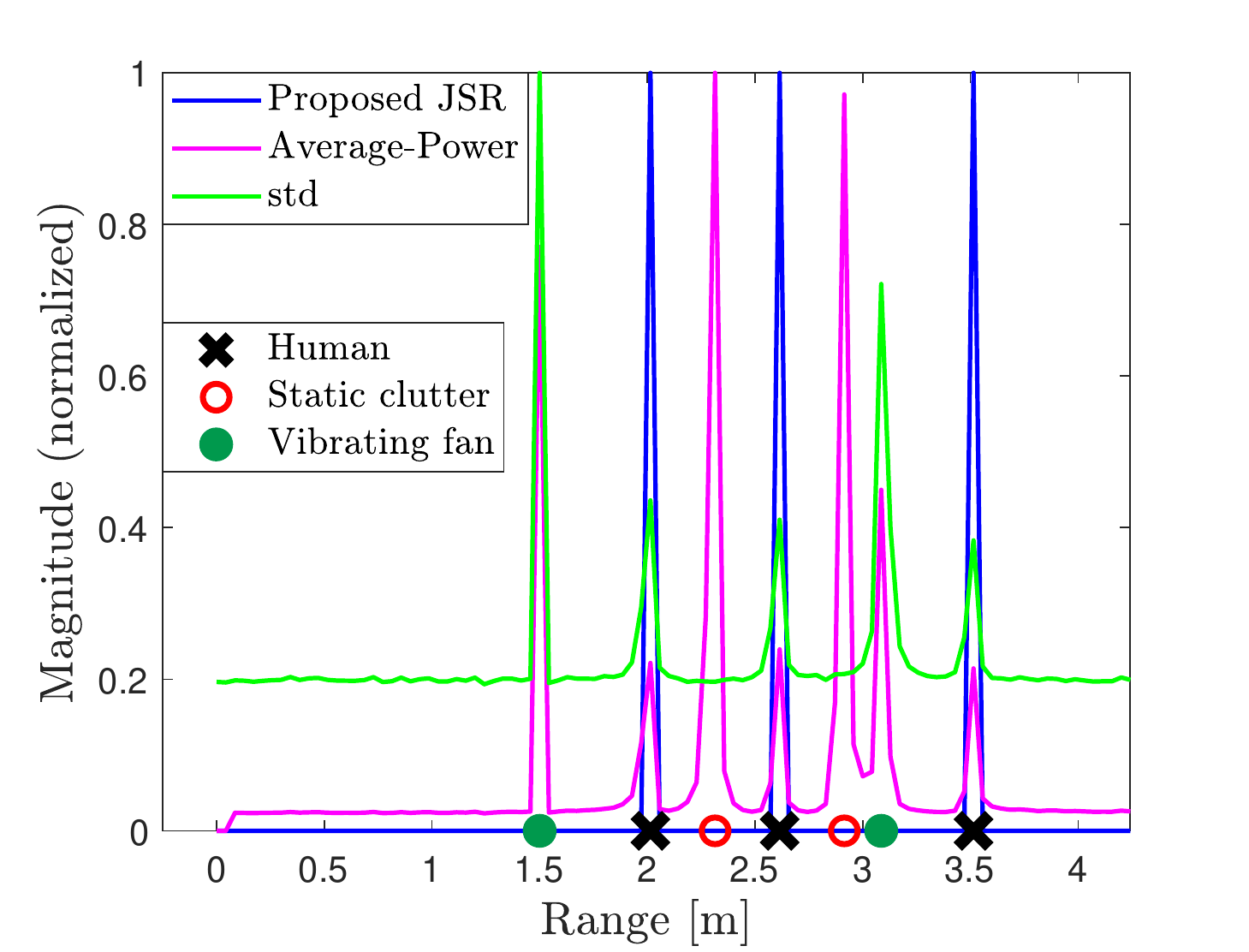}}
\end{center} 
\caption{Multiple people localization by the setup of Table \ref{table:local_exp}, for $\textrm{SNR}=0$ [dB]. \textbf{(a)} The \textit{Range vs. Slow-Time} map via the magnitudes of ${{{{\widehat{{\bf{X}}}}_{M}}}}$. The row-wise intensities correspond to the DC component and the reflections of Table \ref{table:local_exp}'s objects. \textbf{(b)} Several localization techniques. Only the proposed JSR method (\ref{X_rec_JSR}) properly detects the humans in the given scenario. }
\label{fig:MP_localization}
\end{figure*}

By replacing ${\bf{F}}_{\cal{S}}$ in (\ref{X_s_estimate_final}) with ${\bf{F}}_M$, $M=N/2$, the estimator in (\ref{X_s_estimate_final}) evaluates the complete \textit{Range vs. Slow-Time} map by ${{{{{\widehat{{\bf{X}}}}_{M}}}} = \frac{1}{N}{\bf{F}}_{M}{\bf{Y}}}$ and adjusting the \textit{fast-time} axis by (\ref{fm_dm}). Fig. \ref{range_vs_slow_time} depicts the \textit{Range vs. Slow-Time} map via the magnitudes of ${{{{\widehat{{\bf{X}}}}_{M}}}}$, for $\textrm{SNR}=0$ [dB]. The visible row-wise intensities correspond to the DC component as well as the reflections from Table \ref{table:local_exp}'s objects. We note that although the standard practice is to use this map for localizing the humans (Fig. \ref{block_diagram}: block 2.c, works \cite{alizadeh2019remote,mercuri2019vital,sacco2020fmcw}), the proposed method does not.

One can observe in Fig. \ref{localization} that the proposed JSR method indicates the correct locations of the humans compared to the intensity-based \textit{Maximum Average Power} method \cite{alizadeh2019remote} and the \textit{std} approach \cite{sacco2020fmcw}. Notice that the former inadvertently selects the strong reflections obtained from statics clutters, whereas the latter seeks for highly oscillating objects and thus incorrectly selects the vibrating fans over humans. In contrast to the compared localization techniques, the proposed method exploits both characteristics of human-typical vital frequencies and prior knowledge of the sparse structure of ${\tilde{\bf{X}}}$.

\subsection{NCVSM vs. SNR}
Given a successful human localization from the previous study, we examined the performance of VSDR (\ref{VSDR_a_k_rec}) for NCVSM, and compared it to other state-of-the-art techniques using data of $30$ individuals from the resting scenario in \cite{schellenberger2020dataset}.

To ensure a fair comparison of various techniques, we applied the first iteration of Algorithm $1$ assuming that all considered humans were successfully identified. Then, we only compared the last step ($7$) of the algorithm, in which the human vital signs are estimated given the extracted matrix ${\bf{V}}$. To facilitate the analysis, we study here the thoracic vibration ${\bf{v}}_2$ (\ref{V_final}) of human $\#2$ from Table \ref{table:local_exp}, using impedance data from all $30$ participants in \cite{schellenberger2020dataset}. Similar findings are obtained for the remaining vibrations, thus they are not reported. In addition, the frequency bands of respiration and heartbeat were set for all methods to ${B^{\left( R \right)}}=\left[0.1\hspace{0.2cm}0.4 \right]$ [Hz] and
${B^{\left( H \right)}}=\left[0.78\hspace{0.2cm}1.67 \right]$ [Hz], respectively, corresponding to a normal resting state. Note that all the procedures and parameter settings that preceded this step are the same among all methods.

Our proposed VSDR was first compared to the method detailed in \cite{adib2015smart} for estimating RR and HR given the phase of an FMCW signal, called here Phase-Reg. Moreover, VSDR was compared to a FFT-based peak selection in each frequency band, with zero-padding (FFT w/ ZP) \cite{turppa2020vital} and without (FFT w/o ZP) \cite{mercuri2019vital,sacco2020fmcw,adib2015smart,alizadeh2019remote,antolinos2020cardiopulmonary}. The padding of FFT w/ ZP was set to fit a $60$-second time window corresponding to frequency resolution of $1$ [bpm]. All methods were implemented using MATLAB. As to the reference data, we found that it is common to estimate the RR and HR of the reference signals via the DFT spectrum \cite{adib2015smart,turppa2020vital,mercuri2019vital,sacco2020fmcw,alizadeh2019remote}. Assuming that both the ECG and impedance references are noise-free, we padded them similarly to FFT w/ ZP, for optimal results.

To evaluate the monitoring accuracy, we used several statistical metrics, based on the RR and HR estimates of the compared methods w.r.t. those of the references. 1. Success-Rate, defined here as the percentage of time in which the estimate was different from the reference output by less than $2$ [bpm]. 2. Pearson Correlation Coefficient (PCC) \cite{fisher1992statistical},  showing results in the range [$0$ $1$]. 3. Mean-Absolute Error (MAE), and 4. Root-Mean-Square Error (RMSE). 

The Success-Rate can predict the percentage of time an estimation error greater than $2$ [bpm] is expected, 
the PCC measures the linear correlation between the two data sets, and while in the MAE metric, each error contributes in proportion to its absolute value, the RMSE is a well-known metric that emphasizes the occurrence of coarse errors. In this study, we investigate various SNR cases, each of which involves monitoring data of $30$ individuals from \cite{schellenberger2020dataset}. Hence, the performance score produced for each metric, given SNR, is taken as the median across all $30$ participants. We note that $10$ minutes of monitoring starting at $T_{\textrm{win}}=30$ [s], with output obtained at each $T_{\textrm{int}}=0.05$ [s] brings to $11400$ estimates for comparison to the references, for each participant.

\begin{figure*}[htbp!]  
\begin{center}
\vspace{-0.3cm}\hspace{-0.35cm}\subfigure[]{\label{fig:v_2}\includegraphics[width=0.9\textwidth]{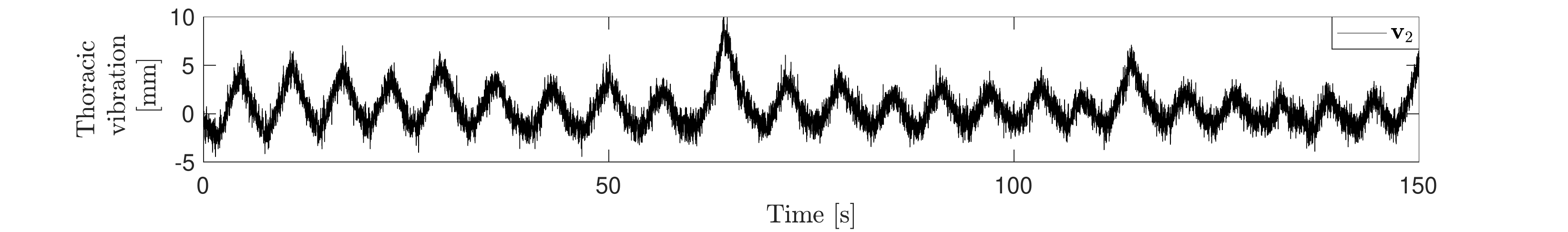}}\\\vspace{-0.3cm}
\subfigure[]{\label{fig:mon_FFT_w_ZP}\includegraphics[width=0.39\textwidth]{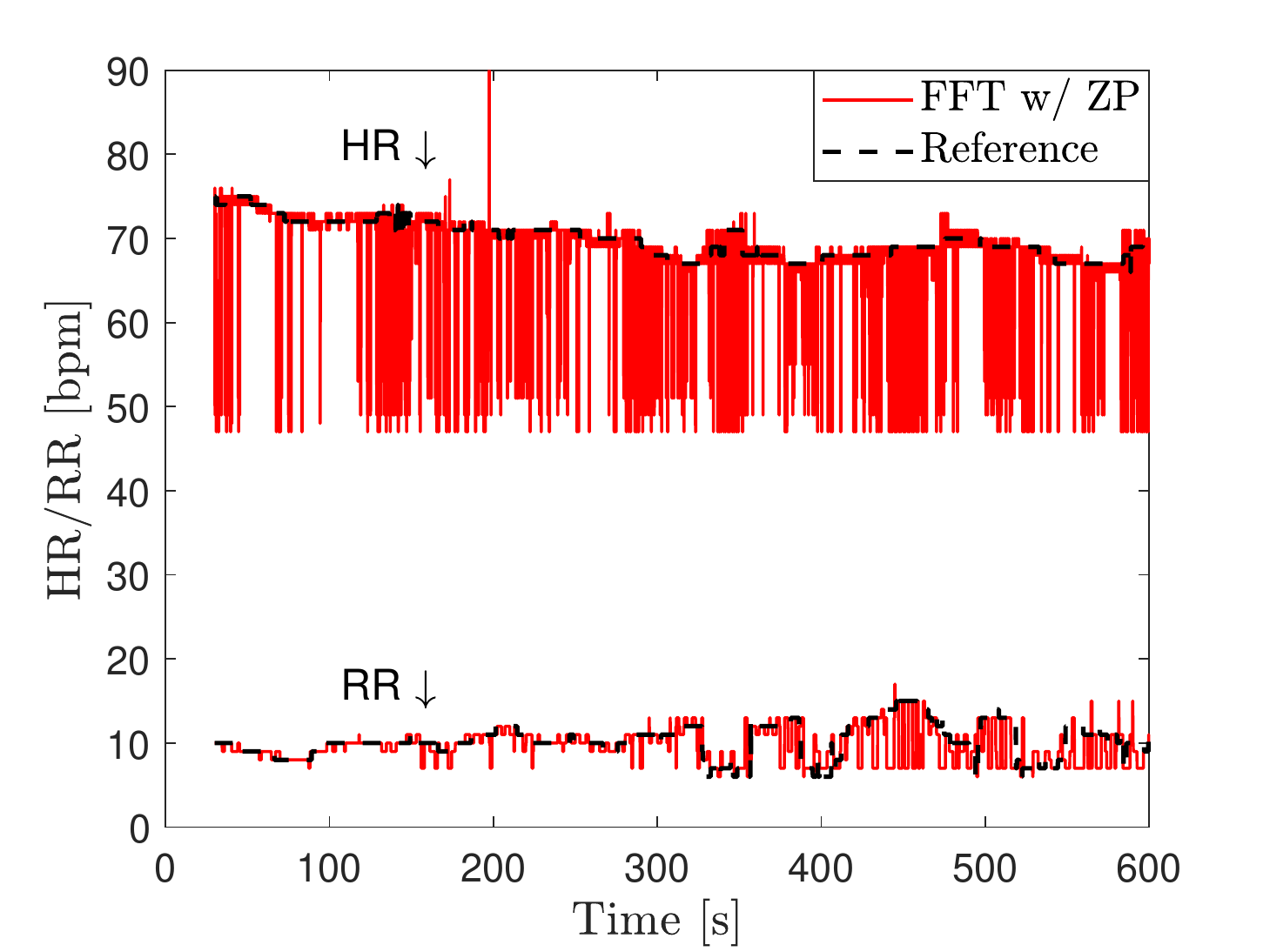}}
\subfigure[]{\label{fig:mon_FFT_wo_ZP}\includegraphics[width=0.39\textwidth]{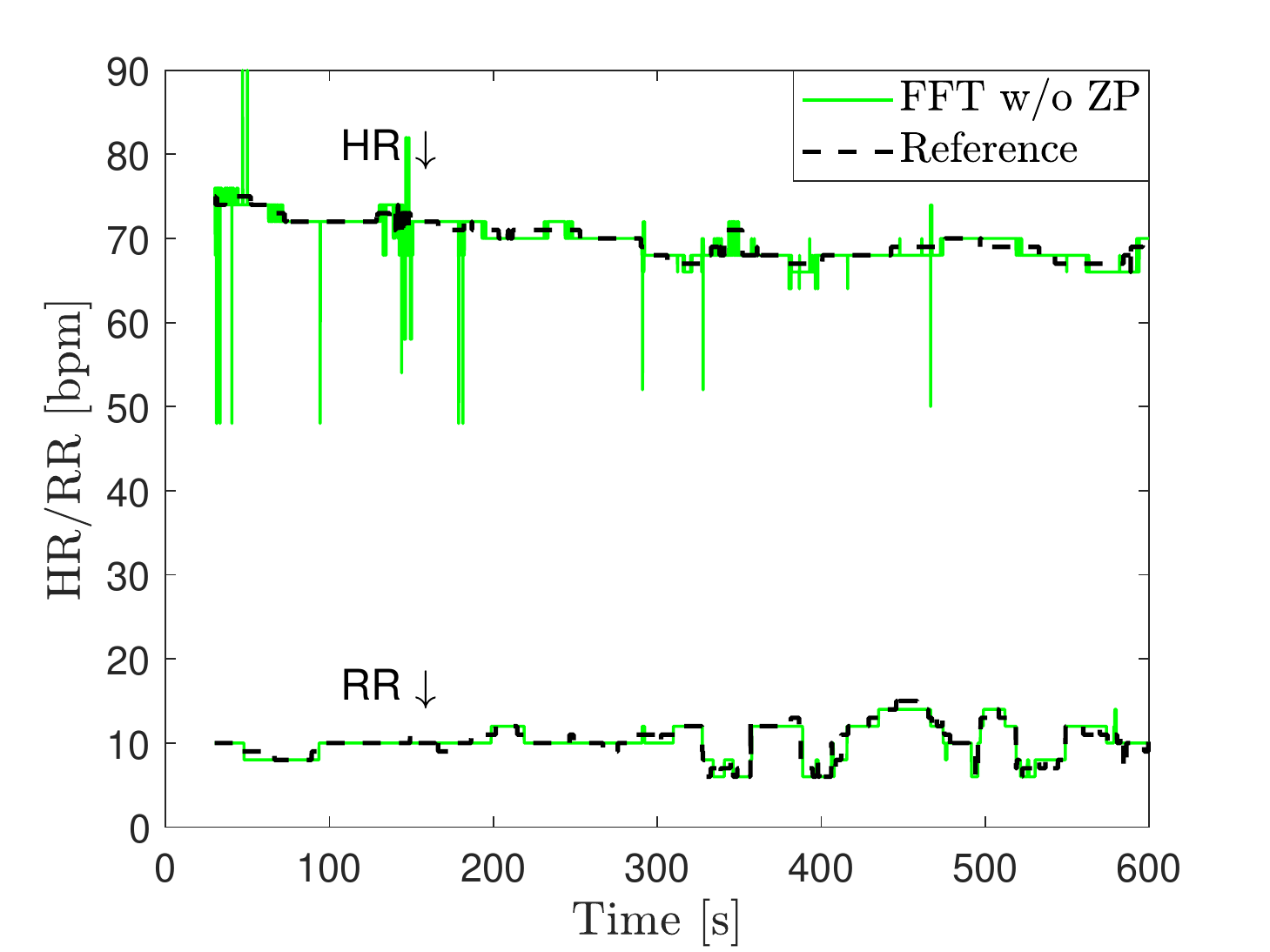}}\\\vspace{-0.3cm}
\subfigure[]{\label{fig:mon_Phase_Reg}\includegraphics[width=0.39\textwidth]{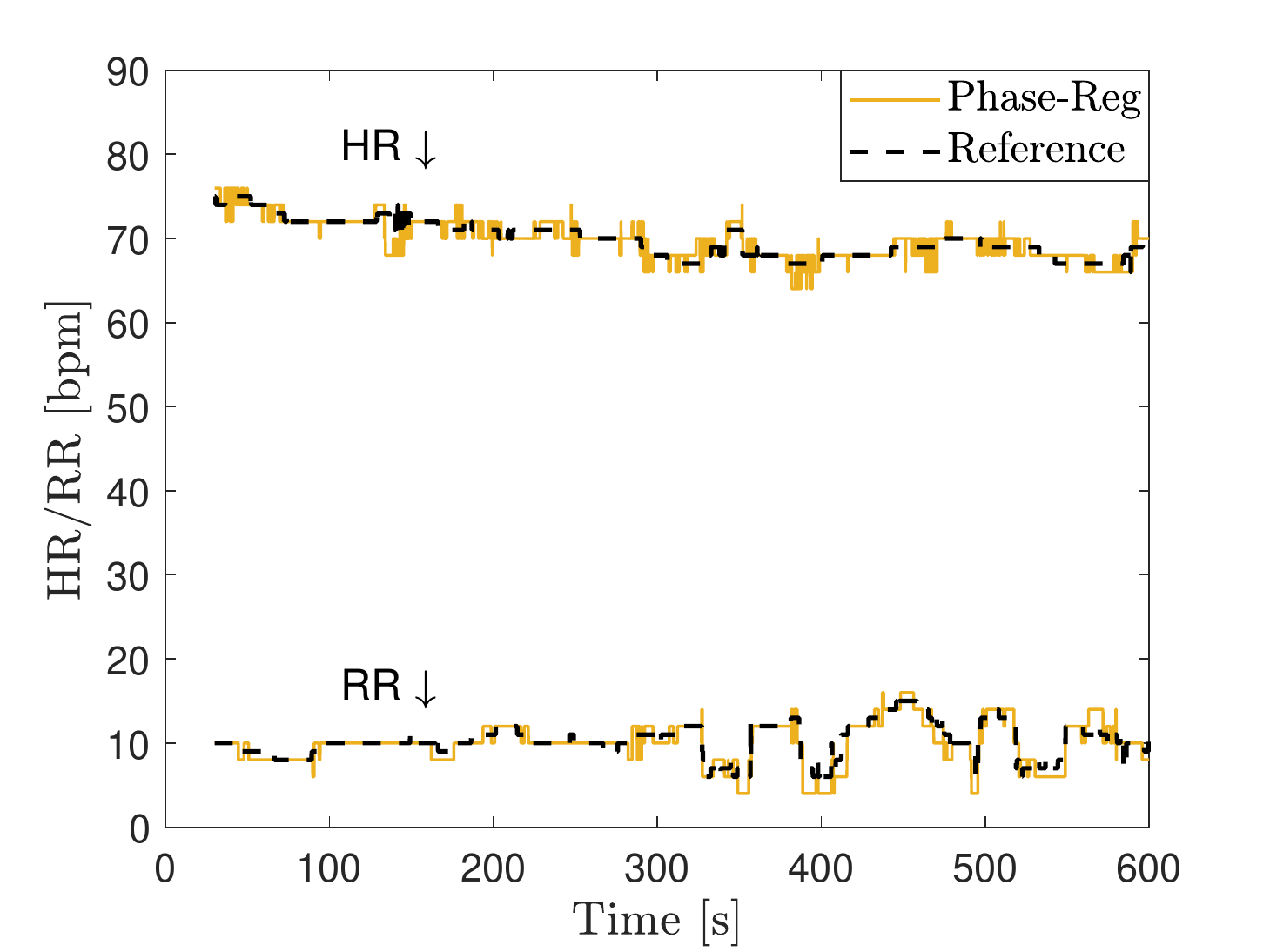}}
\subfigure[]{\label{fig:mon_VSDR}\includegraphics[width=0.39\textwidth]{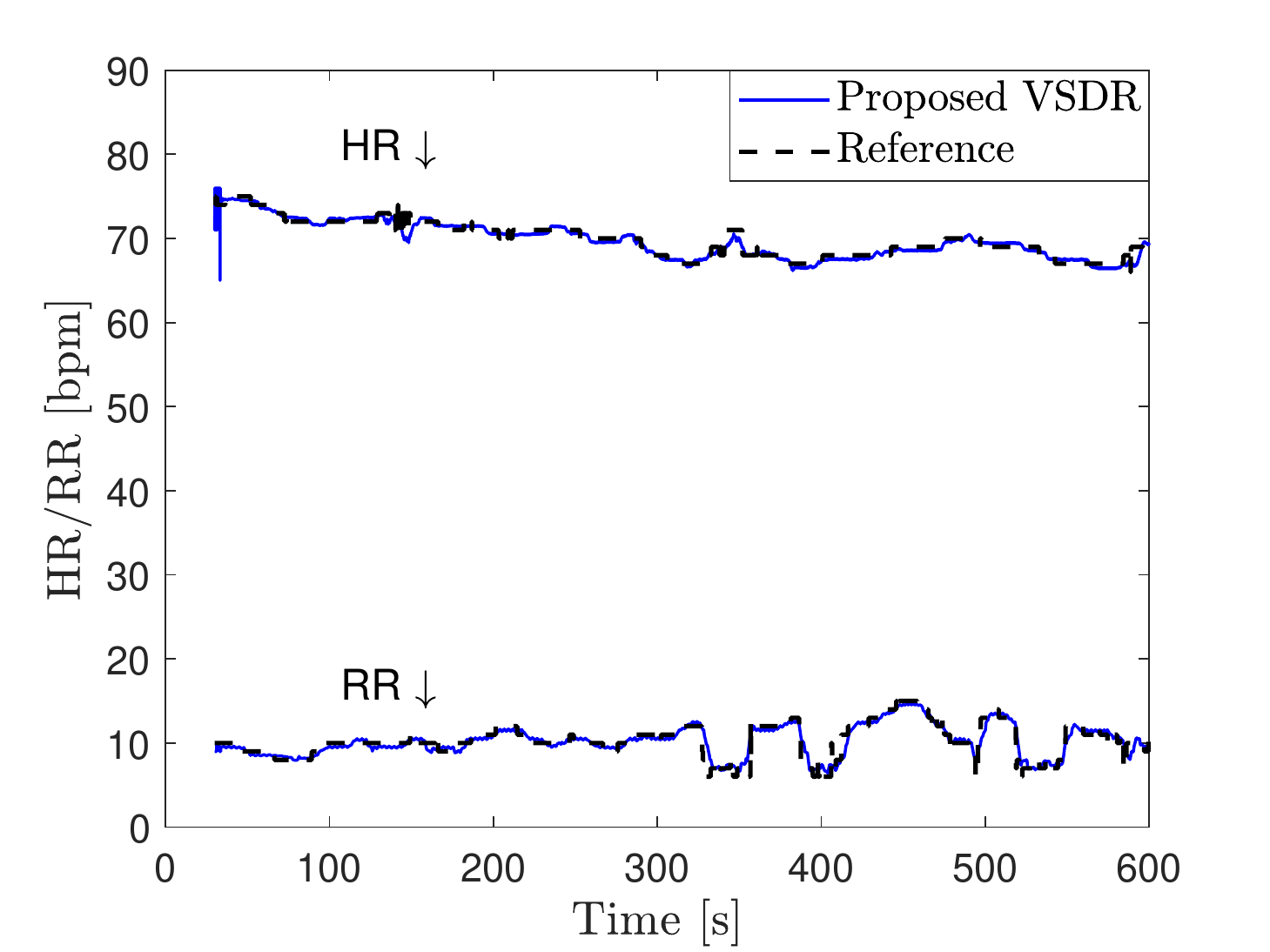}}
\end{center} 
\caption{NCVSM based on subject GDN0004's data \cite{schellenberger2020dataset} for $\textrm{SNR}=1$ [dB], compared to the references. \textbf{(a)} Extracted thoracic vibration ${\bf{v}}_2$. \textbf{(b)} FFT w/ ZP. \textbf{(c)} FFT w/o ZP. \textbf{(d)} Phase-Reg. \textbf{(e)} Proposed VSDR.}   
\label{fig:Monitoring}
\end{figure*}

Fig. \ref{fig:v_2} depicts the extracted thoracic vibration pattern ${\bf{v}}_2$ corresponding to subject GDN0004 \cite{schellenberger2020dataset} for the case of $\textrm{SNR}=1$ [dB]. Figures \ref{fig:mon_FFT_w_ZP}-\ref{fig:mon_VSDR} show NCVSM compared to the references, by VSDR and the other examined techniques, given the extracted vibration ${\bf{v}}_2$ at each $T_{\textrm{int}}$. It can be seen how both the HR and RR estimates by the proposed VSDR show great resemblance to those of the reference, compared to the other techniques in which the noisy setup impairs the evaluations. 

\begin{figure*}[htbp!]  
\begin{center}
\vspace{-0.5cm}
\subfigure[]{\label{}\includegraphics[width=0.24\textwidth]{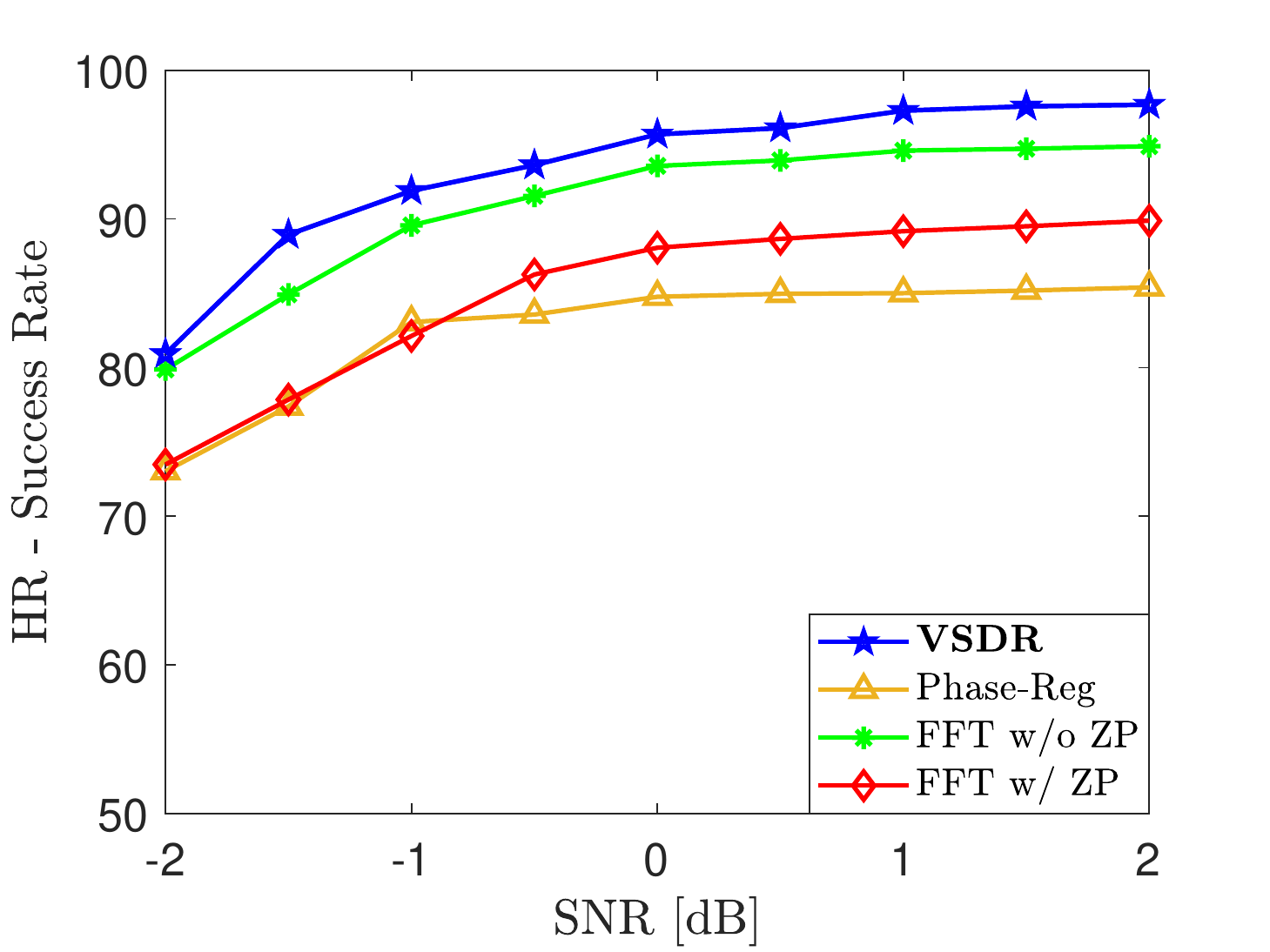}}
\subfigure[]{\label{}\includegraphics[width=0.24\textwidth]{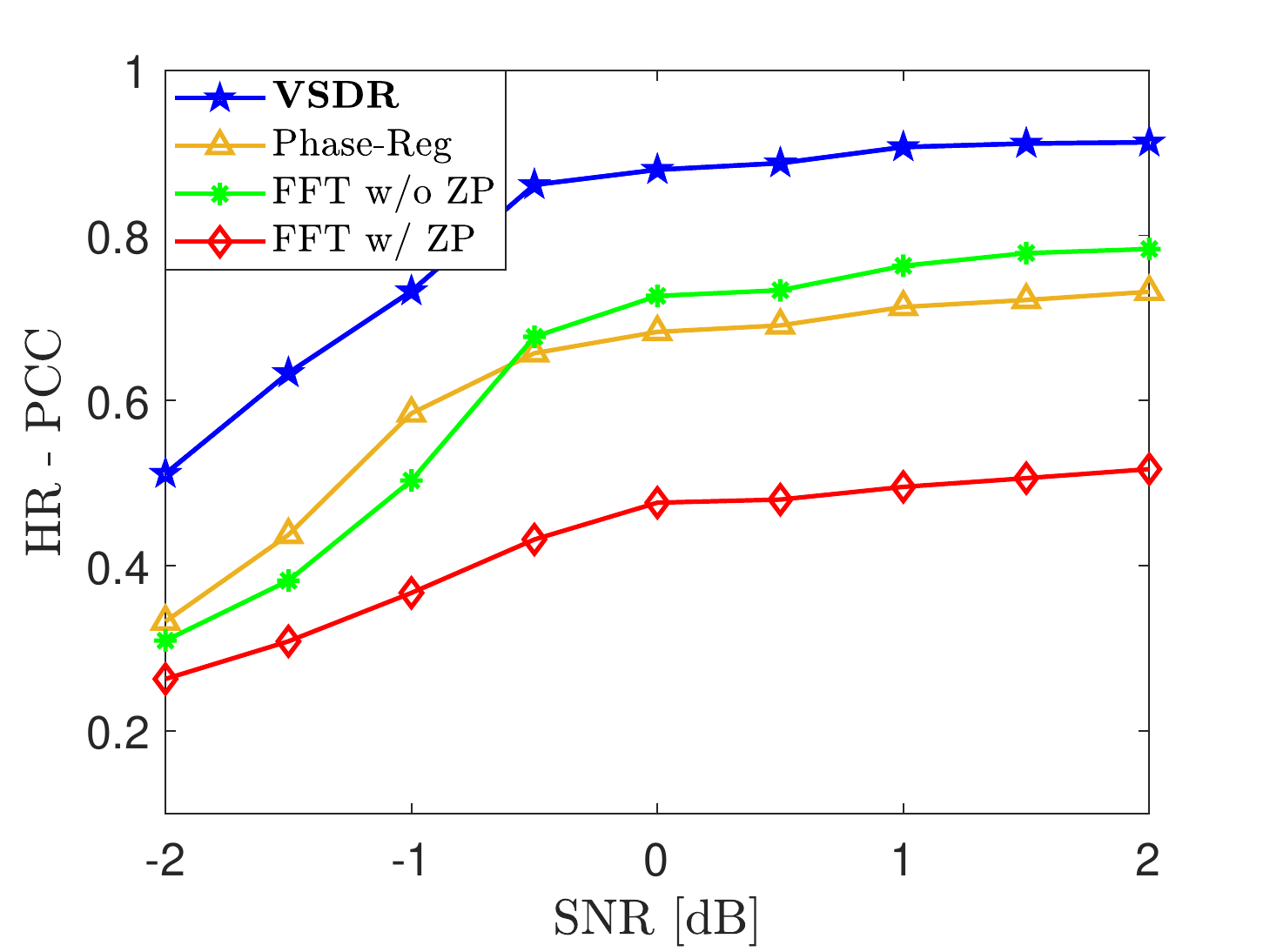}}
\subfigure[]{\label{}\includegraphics[width=0.24\textwidth]{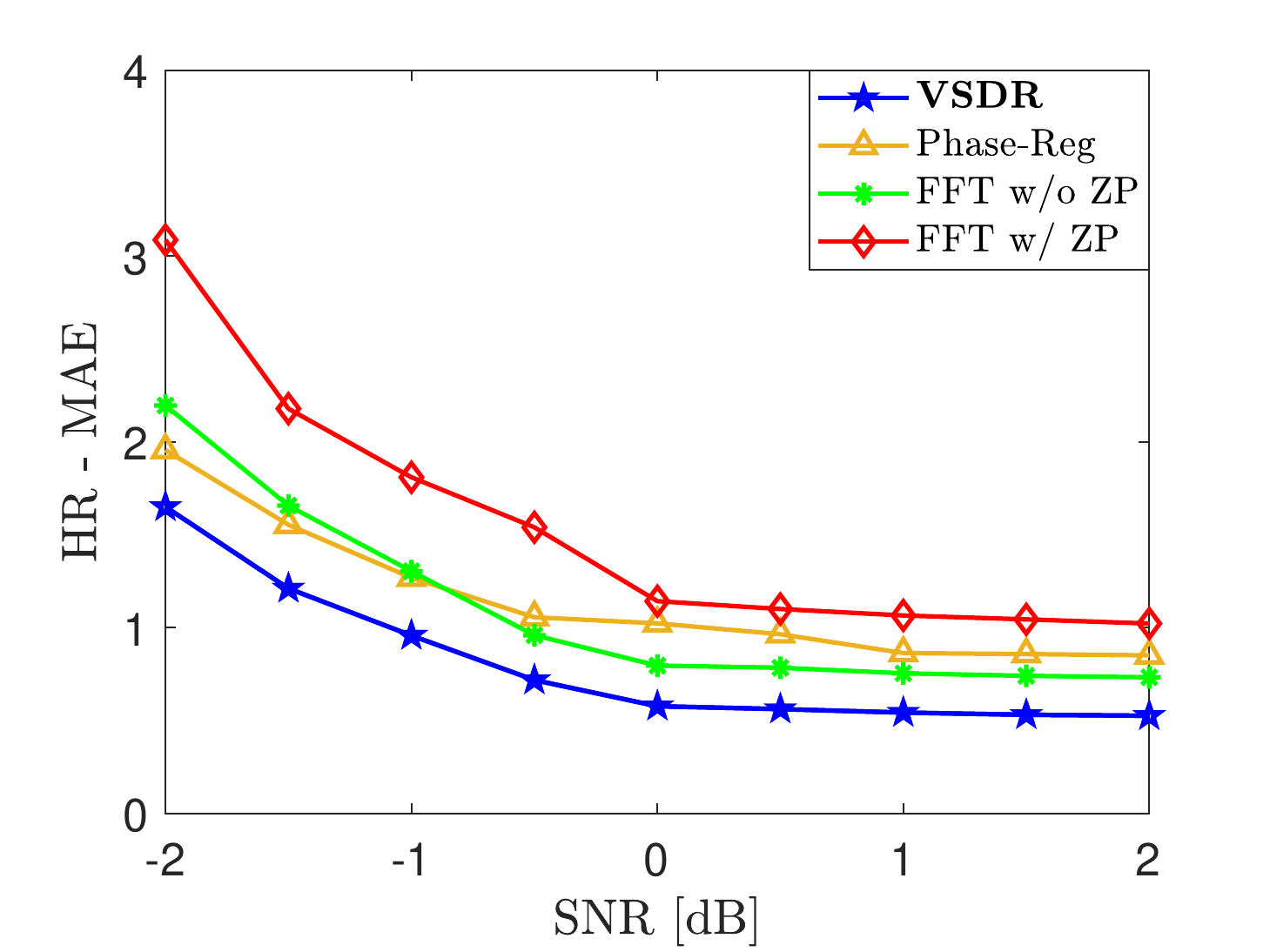}}
\subfigure[]{\label{}\includegraphics[width=0.24\textwidth]{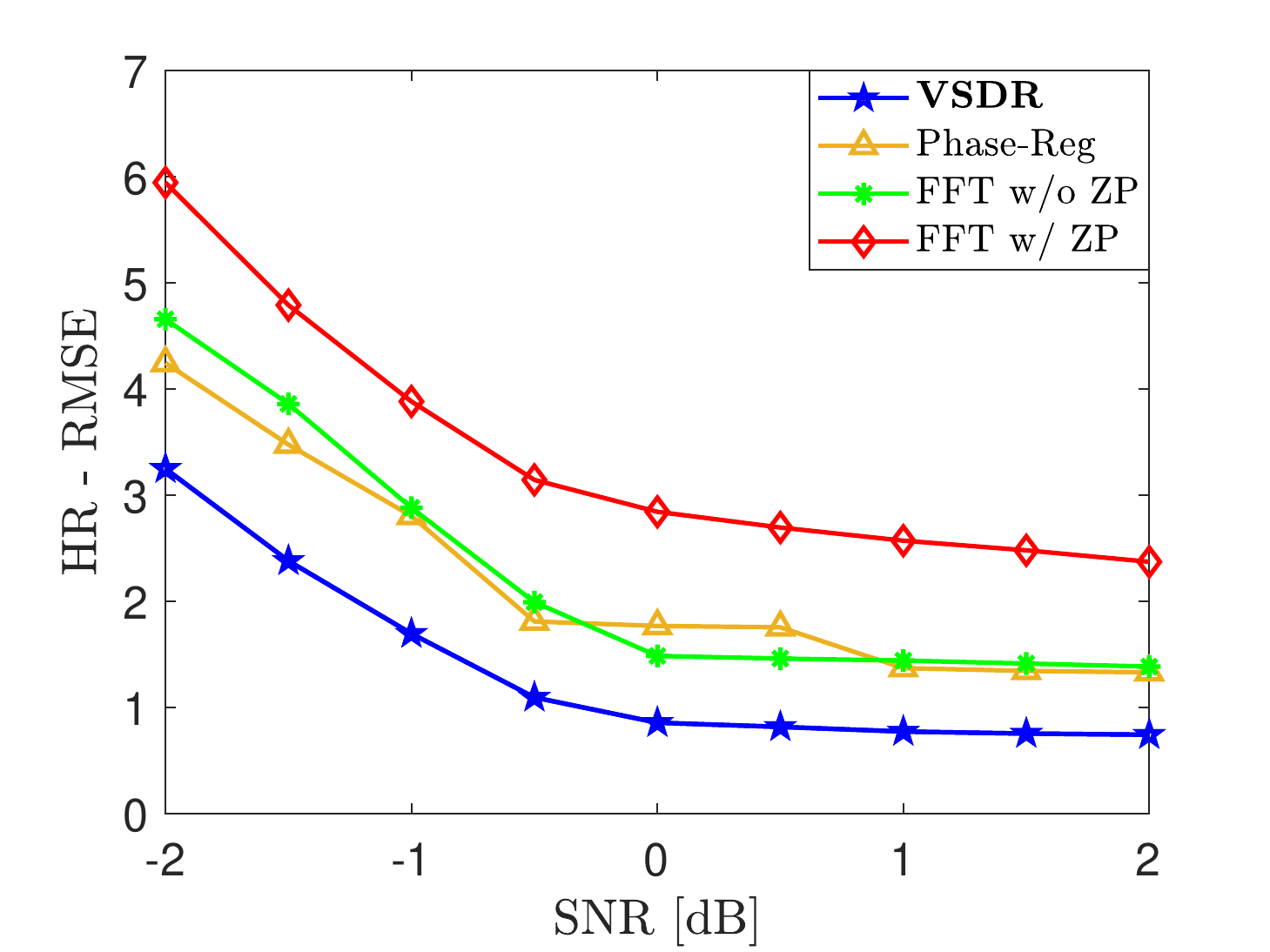}}\\\vspace{-0.3cm}
\subfigure[]{\label{}\includegraphics[width=0.24\textwidth]{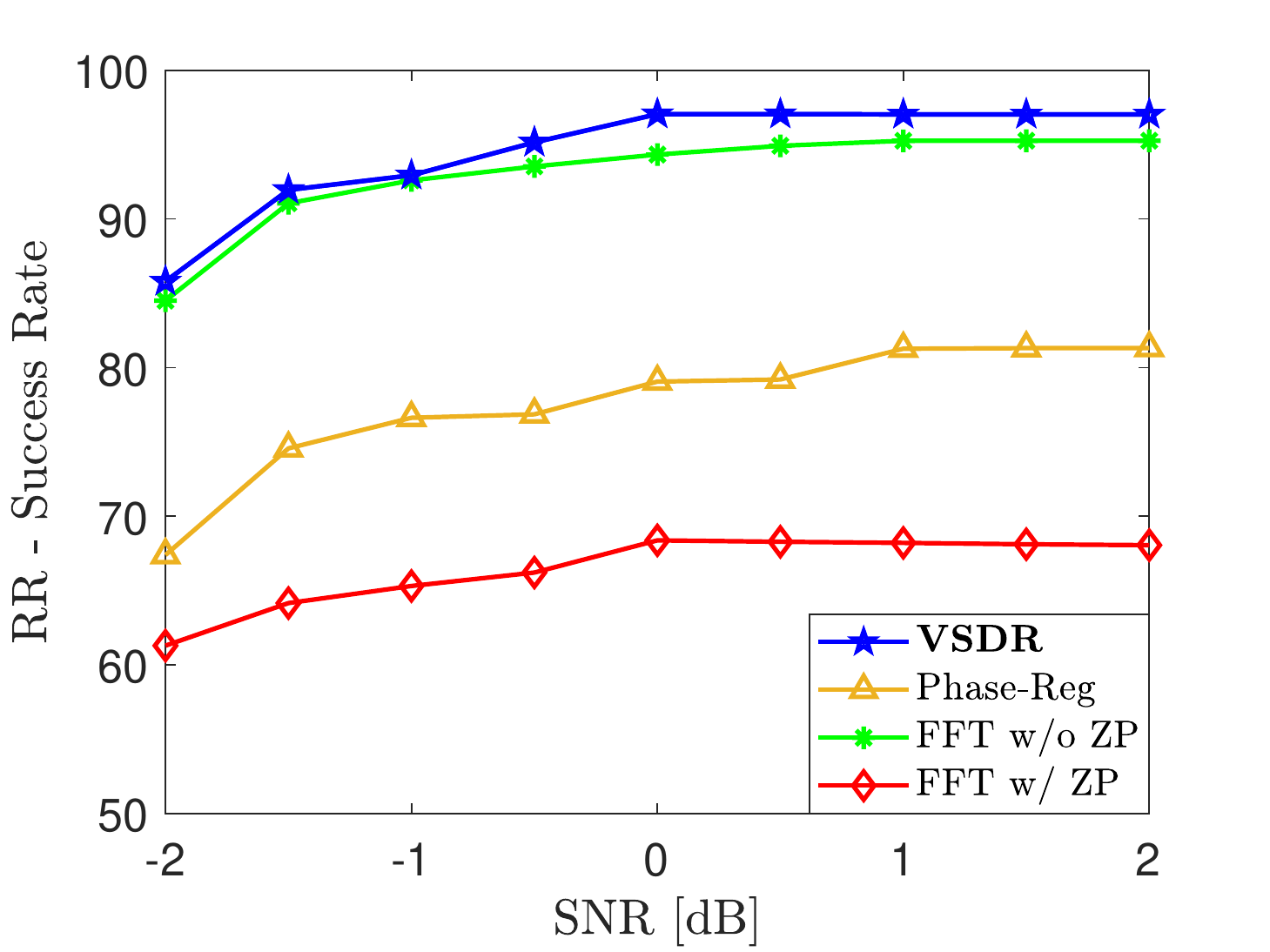}}
\subfigure[]{\label{}\includegraphics[width=0.24\textwidth]{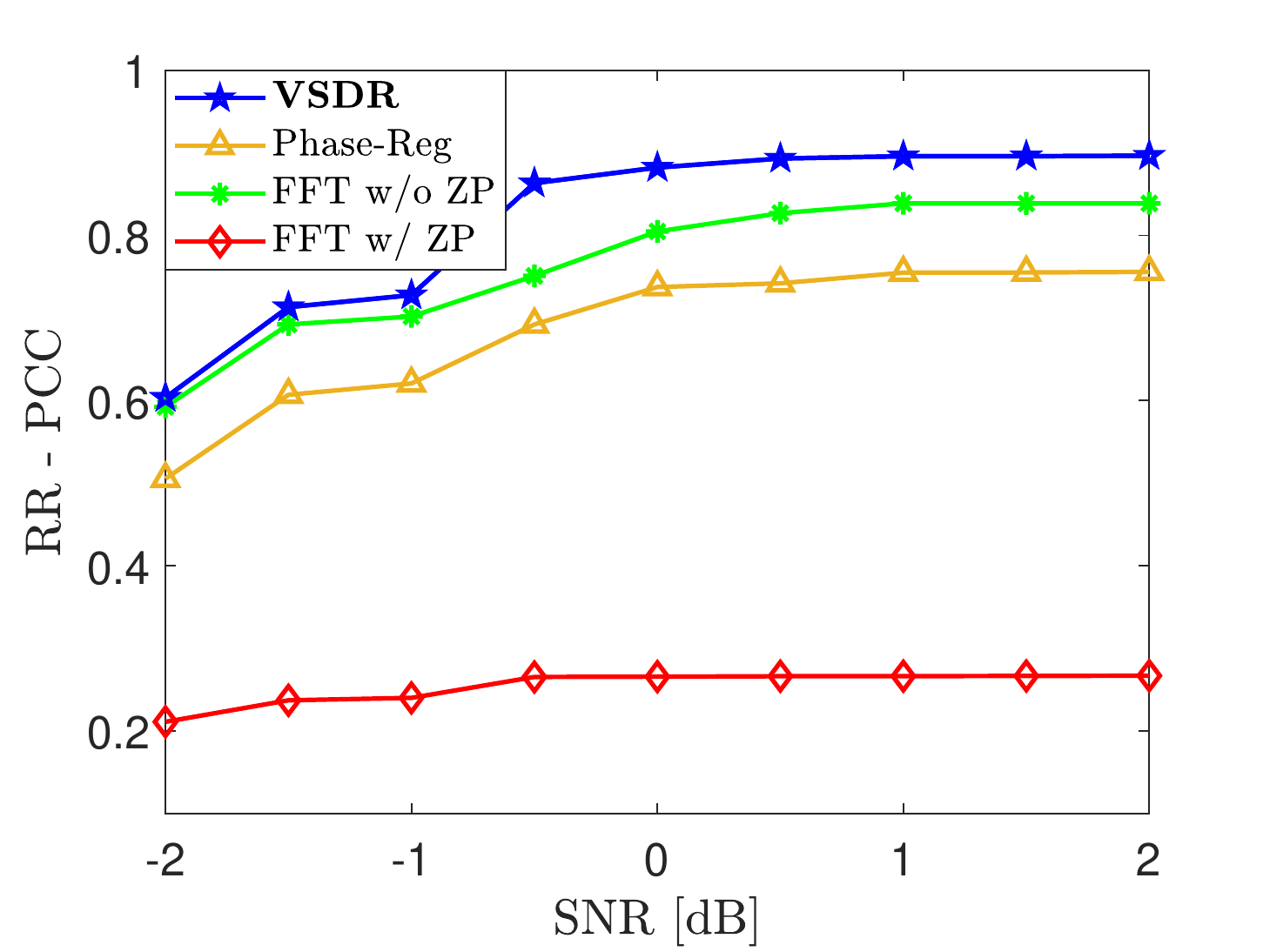}}
\subfigure[]{\label{}\includegraphics[width=0.24\textwidth]{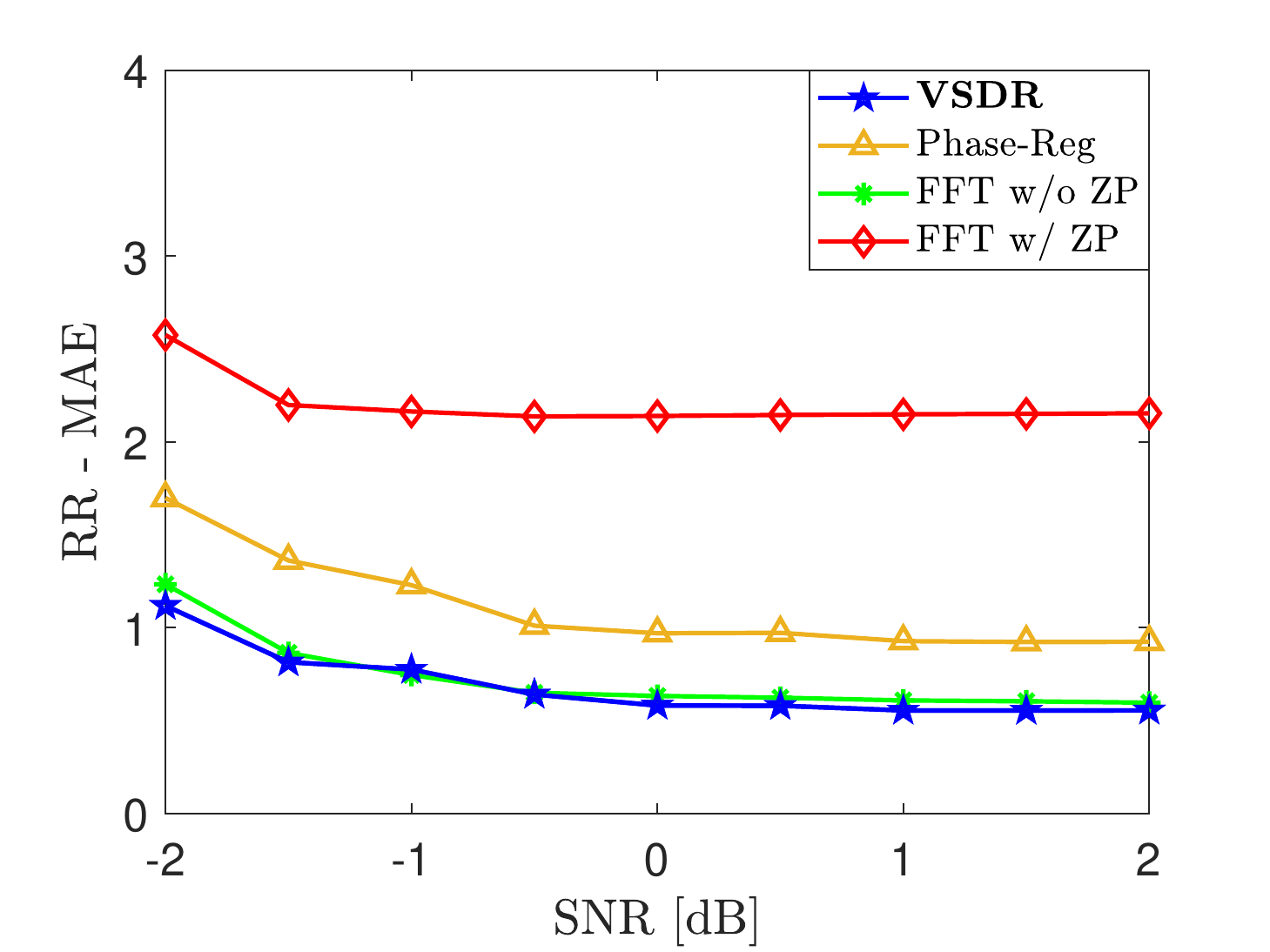}}
\subfigure[]{\label{}\includegraphics[width=0.24\textwidth]{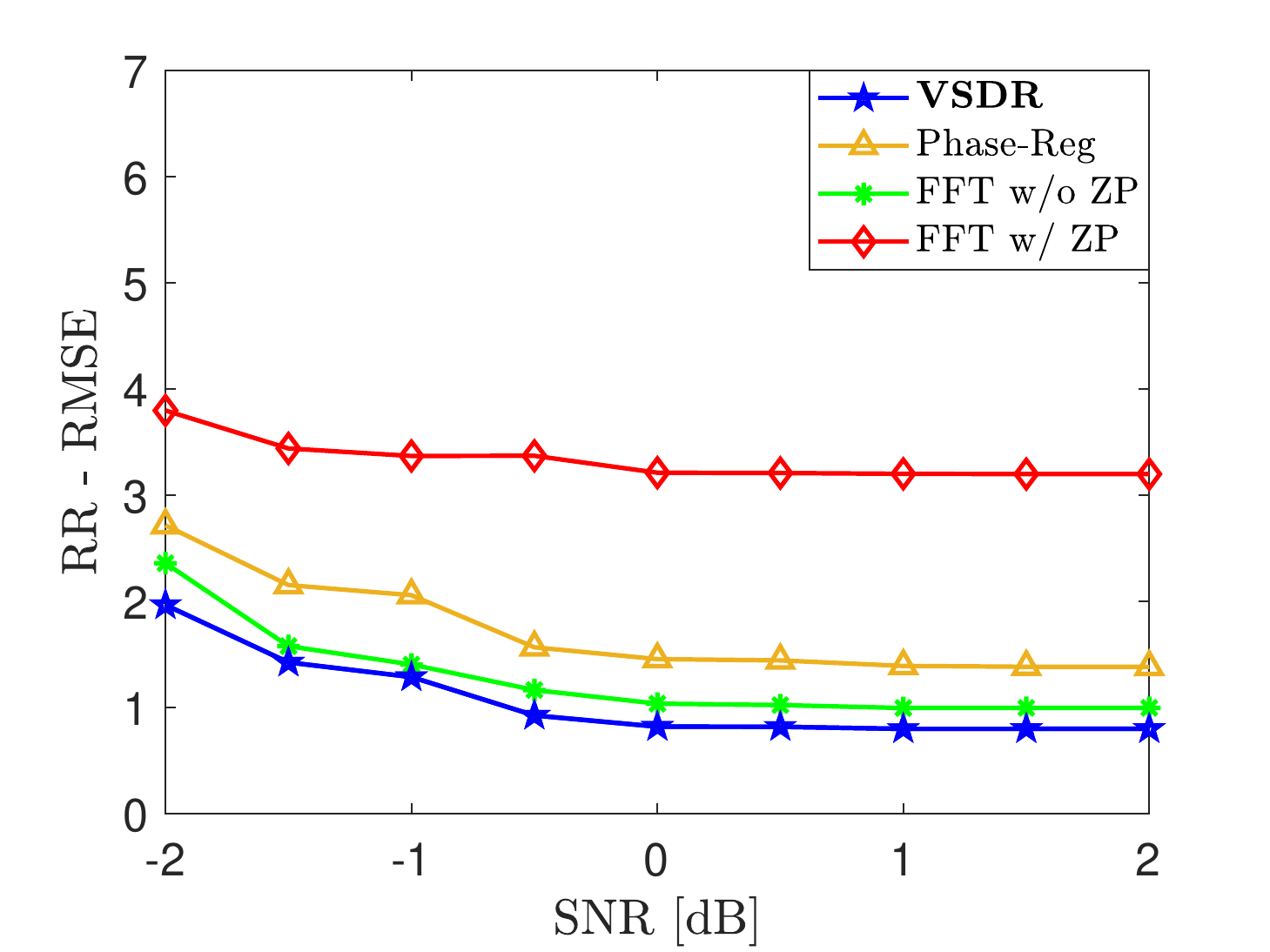}}
\end{center} 
\caption{HR and RR monitoring performance vs. $\textrm{SNR} \in [-2,\hspace{0.1cm}2]$. \textbf{Top:} \textbf{(a)} HR Success Rate. \textbf{(b)} HR PCC. \textbf{(c)} HR MAE. \textbf{(d)} HR RMSE.\\\textbf{Bottom:} \textbf{(e)} RR Success Rate. \textbf{(f)} RR PCC. \textbf{(g)} RR MAE. \textbf{(h)} RR RMSE.}   
\label{fig:Performance_curve}
\end{figure*}

Fig. \ref{fig:Performance_curve} shows the Success-Rate, PCC, MAE and RMSE for both HR and RR estimation by all examined methods, as a function of the SNR. One sees that VSDR outperforms the other compared methods in all $4$ metrics, for every SNR value. Also, note the considerable difference in performance in the more challenging task of HR monitoring due to the weak heartbeat signature, in favor of the proposed approach. As for the RR monitoring case, the small difference between VSDR and the popular FFT w/o ZP can be explained by the relative dominance of the respiratory signal that facilitates its detection, bringing to fine results in both techniques. Detailed median accuracy scores for the noisy case of $\textrm{SNR}=0$ [dB] can be found in tables \ref{fig:HR_performance_table} and \ref{fig:RR_performance_table}.

The performance advantage is a consequence of the property that, unlike all other compared methods for NCVSM given a thoracic vibration, the proposed VSDR employs a frequency search over high resolution grids corresponding to human-typical cardiopulmonary frequencies via a dictionary-based approach (\ref{VSDR_a_k_rec}).

\newpage

\begin{table}[h!]
\centering
\begin{tabular}{ |c | c|c| c|c| }
\hline
Method & Success-Rate [$\%$] & PCC & MAE & RMSE\\
\hline
FFT w/ ZP & 88.06  & 0.47 & 1.14 & 2.84\\
\hline
FFT w/o ZP & 93.56 & 0.72 & 0.79 & 1.48\\
\hline
Phase-Reg & 84.76 & 0.68 & 1.02 & 1.76\\
\hline
\textbf{Proposed VSDR} & \textbf{95.68 } & \textbf{0.87} & \textbf{0.57} & \textbf{0.85}\\
\hline
\end{tabular}
\caption{HR estimation median accuracy for the noisy case of $\textrm{SNR}=0$ [dB].}
\label{fig:HR_performance_table}
\end{table}

\begin{table}[h!]
\centering
\begin{tabular}{ |c | c|c| c|c| }
\hline
Method & Success-Rate [$\%$] & PCC & MAE & RMSE\\
\hline
FFT w/ ZP & 68.37 & 0.26 & 2.13 & 3.20\\
\hline
FFT w/o ZP & 94.32 & 0.80 & 0.63 & 1.03\\
\hline
Phase-Reg & 79.05 & 0.73 & 0.97 & 1.45\\
\hline
\textbf{Proposed VSDR} & \textbf{97.04} & \textbf{0.88} & \textbf{0.58} & \textbf{0.81}\\
\hline
\end{tabular}
\caption{RR estimation median accuracy for the noisy case of $\textrm{SNR}=0$ [dB].}
\label{fig:RR_performance_table}
\end{table}

\section{Conclusion}
\label{Conc}
In this paper, an extended FMCW signal model for NCVSM of multiple people was derived, allowing to interpret a realistic noisy environment containing multiple objects. Based on this model, we presented a JSR approach that accurately localizes multiple people in a clutter-rich scenario, based on the sparse composition of the input data. We then developed the VSDR method, which performs accurate NCVSM given human thoracic vibrations, by leveraging human-typical cardiopulmonary characteristics using a dictionary based approach. The robustness of the proposed VSDR is reflected in superior performance results using data from $30$ monitored individuals, outperforming state-of-the-art alternatives using multiple statistical criteria.

\section*{Acknowledgement}
The authors would like to thank Daniel Khodyrker for his assistance with the numerical study. 





\end{document}